\documentclass[iop, apj, revtex4]{emulateapj}
\usepackage{natbib}
\usepackage{url}
\usepackage{amsmath} 
\usepackage{appendix}
\usepackage{hyperref}

\begin{document}

\title{Hubble Space Telescope Observations of Accretion-Induced Star Formation in the Tadpole Galaxy Kiso 5639}

\author{Debra Meloy Elmegreen\altaffilmark{1},
Bruce G. Elmegreen\altaffilmark{2}, Jorge S\'anchez Almeida\altaffilmark{3}, Casiana
Mu\~noz-Tu\~n\'on\altaffilmark{3}, Jairo Mendez-Abreu\altaffilmark{4}, John S.
Gallagher\altaffilmark{5}, Marc Rafelski\altaffilmark{6}, Mercedes Filho\altaffilmark{3},
Daniel Ceverino\altaffilmark{7}} \altaffiltext{1}{Vassar College, Dept. of Physics and
Astronomy, Poughkeepsie, NY 12604} \altaffiltext{2}{IBM Research Division, T.J. Watson Research
Center, Yorktown Hts., NY 10598} \altaffiltext{3}{Instituto de Astrof\'isica de Canarias, C/
via L\'actea, s/n, 38205, La Laguna, Tenerife, Spain, and Departamento de Astrof\'isica,
Universidad de La Laguna} \altaffiltext{4}{School of Physics \& Astronomy, North Haugh, St
Andrews, KY169SS, UK} \altaffiltext{5}{Dept. of Astronomy, Univ. of Wisconsin-Madison, Madison,
WI 53706} \altaffiltext{6}{NASA Postdoctoral Program Fellow, Astrophysics Science Division,
Goddard Space Flight Center, Code 665, Greenbelt, MD 20771} \altaffiltext{7}{Zentrum f\"ur
Astronomie der Universit\"at Heidelberg, Institut f\"ur Theoretische Astrophysik,
Albert-Ueberle-Str. 2, 69120 Heidelberg, Germany}

\begin{abstract}
The tadpole galaxy Kiso 5639 has a slowly rotating disk with a drop in metallicity at its
star-forming head, suggesting that star formation was triggered by the accretion of
metal-poor gas. We present multi-wavelength HST WFC3 images of $UV$ through $I$ band plus
H$\alpha$ to search for peripheral emission and determine the properties of various
regions. The head has a mass in young stars of $\sim 10^6\;M_{\odot}$ and an
ionization rate of $6.4\times10^{51}$ s$^{-1}$, equivalent to $\sim2100$ O9-type stars.
There are four older star-forming regions in the tail, and an underlying disk with a
photometric age of $\sim1$ Gyr. The mass distribution function of 61 star clusters is a
power law with a slope of $-1.73\pm0.51$. Fourteen young clusters in the head are more
massive than $10^4\;M_\odot$, suggesting a clustering fraction of 30\%-45\%. Wispy
filaments of H$\alpha$ emission and young stars extend away from the galaxy. Shells and
holes in the head H\textsc{ii} region could be from winds and supernovae. Gravity
from the disk should limit the expansion of the H\textsc{ii} region, although hot
gas might escape through the holes. The star formation surface density
determined from H$\alpha$ in the head is compared to that expected from likely
pre-existing and accreted gas. Unless the surface density of the accreted gas is a factor
of $\sim3$ or more larger than what was in the galaxy before, the star formation rate has
to exceed the usual Kennicutt-Schmidt rate by a factor of $\ge5$.

\end{abstract} 

\section{Introduction}
Tadpole galaxies are characterized by a bright star-forming head and an elongated tail with
weak star formation. They are sometimes called cometary \citep{mark,loose,cairos} and are
designated sub-class iI,c of the Blue Compact Dwarf (BCD) type by \cite{noeske00}. Other
local tadpoles were studied by \cite{kniazev}, \cite{paz}, \cite{pap}, \cite{elm12}, and
others. They are rare in the local universe, where only 0.2$\%$ of the 10,000 galaxies in
the Kiso Survey of UV Bright Galaxies \citep{kiso-all} are tadpoles \citep{elm12}, but they are common, 75\% \citep{morales}, among local
extremely metal poor (XMP) galaxies, which have metallicities less than 0.1 solar.

The association between tadpole shapes and BCD or low metallicity galaxies suggests an
element of youthfulness.  Indeed, tadpoles are relatively common in the early Universe,
amounting to some 10\% of all galaxies larger than 10 pixels in diameter in the Hubble Space Telescope Ultra Deep Field
 \citep{elm05}, and a similar fraction in other deep
fields \citep{vdb,straughn,wind}. XMP galaxies are also gas-rich, giving them a low average
H\textsc{i} star formation efficiency ($=$ star formation surface density/H\textsc{i}
surface density), but they have relatively few stars, so their specific star formation
rates (SFR/stellar mass) are high, comparable to those in high redshift galaxies
\citep{filho15}.

\begin{deluxetable*}{llccccc}[h!]
\tabletypesize{\scriptsize}\tablecolumns{5} \tablewidth{0pt} \tablecaption{HST WFC3 Observations}
\tablehead{\colhead{RA}&\colhead{Dec}&\colhead {filter}&\colhead{Exposure time} & \colhead{Date}&  \\
\colhead{(J2000)}&\colhead{J2000}&\colhead{}&\colhead{(sec)}&\colhead{}}
\startdata
11 41 09.410  &  +32 25 27.21  &   F225W    &   2508.0    &       02-16-2015\\
11 41 09.410  &   +32 25 27.21  &  F336W   &  2508.0   & 02-16-2015   \\
11 41 08.702 & +32 25 40.21 & F438W & 1248.0 & 07-02-2015\\
11 41 08.702 & +32 25 40.21 & F547M & 1248.0 & 07-02-2015\\
11 41 08.702 & +32 25 40.21 & F606W & 1310.0 & 07-02-2015\\
11 41 08.702 & +32 25 40.21 & F814W & 1310.0 & 07-02-2015\\
11 41 08.640 & +32 25 41.13 & F657N & 2624.0 & 07-02-2015
\label{tab1}
\enddata
\end{deluxetable*}

Single slit spectra of 22 tadpoles or XMP galaxies have been used to determine
metallicities along the major axes and, in a few cases, rotation curves. \cite{jorge13}
obtained spectra of 7 tadpoles with the 2.5m Isaac Newton Telescope and the 2.5m Nordic
Optical Telescope on Tenerife.  Rotation was observed, with a peak speed of several tens of
km s$^{-1}$, as expected for these low mass objects. The metallicities were unusual,
however. As inferred from the N2 index (the ratio [N II]/H$\alpha$), the metallicities were
lowest in the star-forming heads in 6 out of the 7 cases. \cite{jorge14a} determined
metallicities in 7 more dwarf galaxies with the 4.2 m William Hershel Telescope using the
``direct method'' (i.e., [SII] lines $\lambda6717/\lambda6731$ for electron density, [OIII]
lines $(\lambda4959+\lambda5007)/\lambda4363$ for electron temperature of [OIII], and [OII]
lines $\lambda3727/(\lambda7319 + \lambda7330$) for electron temperature of [OII], with a
sum of the oxygen states up to [OIII] to give the oxygen abundance relative to hydrogen).
Of these 7 galaxies, 2 out of 5 with measurable metallicities had low values in the head; the other
2 were ambiguous. Spectra of another 10 XMPs, including tadpoles, were obtained with the
10.4m Gran Telescopio Canarias using the HII-CHI-mistry algorithm \citep[][i.e., minimizing
the $\chi^2$ fit to many lines]{perez14}. In 9 out of 10 cases, the metallicities were
lower in the heads by factors of 3 to 10 \citep{jorge15}.

We interpret these metallicity drops as the result of recent accretion of metal-poor gas
onto part of an otherwise quiescent dwarf galaxy.  Idealized simulations \citep{verbeke14}
as well as cosmological simulations of galaxy formation \citep{ceverino16} show this
effect. A recent review of accretion-fed star formation is discussed in \cite{jorge14b}.
Support for this interpretation comes from the locations of most of the XMPs in low-density
environments, such as voids and sheets in the distribution of galaxies \citep{filho15a}.
They are surrounded by low-metallicity, often-asymmetric H\textsc{i} \citep{filho13}, which
could be the accreting gas reservoir. Motion through this reservoir could cause the
accretion and promote star formation on the leading edge of the dwarf, as might have
happened for 30 Doradus in the Large Magellanic Cloud \citep{mas09}. High-mass late-type
galaxies could also be bursting with star formation in their outer parts from episodic
accretion \citep{huang13}.

One of the best examples of a metallicity drop is in the bright star-forming head of the
local tadpole galaxy Kiso 5639, which is the focus of the present paper. Here we examine
Hubble Space Telescope (HST) Wide Field Camera 3 (WFC3) multiwavelength images of Kiso 5639
to determine the masses and ages of the clusters and large-scale star-forming clumps, and
to search for evidence of accretion on the periphery. Our data are presented in Section
\ref{sect:reduc}, analyses are in Section \ref{sect:analysis}, results are in Section
\ref{sect:results}, and conclusions are in Section \ref{sect:conc}.

\section{Observations and Data Reduction}\label{sect:reduc}

Kiso 5639 (LEDA-36252, KUG 1138+327) is a small emission line galaxy approximately 2.7
kpc in diameter at a galactocentric distance of 24.5 Mpc (NED\footnote{NASA/IPAC
Extragalactic Database, {\url{http://ned.ipac.caltech.edu}}}). It is slowly rotating with
a velocity of 35-40 km s$^{-1}$, giving it a dynamical mass of 1.5$\times 10^8$
$M_{\odot}$ \citep{jorge13}. Its photometric mass is 5$\times 10^7$ $M_{\odot}$ based on
Sloan Digital Sky Survey images \citep{elm12}, and the H\textsc{i} mass is around
$3\times10^8\;M_\odot$ \citep[][see Sect. \ref{gasmass}]{salzer}. HST WFC3/UVIS images of
Kiso 5639 were obtained in February and July 2015 (Proposal 13723) in filters F225W,
F336W, F438W, F547M, F606W, F814W, and F657N (H$\alpha$ + N II), where the parentheses
indicate wide (W), medium (M), and narrow (N) filters with central wavelengths in
nanometers; exposure times and coordinates for the observations are listed in Table 1. At
the assumed distance, the pixel size of 0.0396 arcsec corresponds to 4.7 pc. We placed
the galaxy near the edge of the detector to put it close to the read-out to minimize
charge transfer efficiency (CTE) degradation effects, and adjusted its position with the
F225W and F336W filters to avoid a bright star.

The WFC3/UVIS instrument has degraded over time due to radiation damage, reducing the
CTE, which results in smeared images along the read-out direction
when not corrected \citep{MacKenty:2012}. The science images are therefore CTE corrected using
code provided by STScI\footnote{\url{http://www.stsci.edu/hst/wfc3/tools/cte_tools}}. However,
the dark calibrations also suffer from this degradation, and therefore we required special calibration files
to mask all the hot pixels, and remove a background gradient and splotchy pattern in the
science images. To do so, we created custom super-darks following the method described in \cite{rafelski15},
which use all CTE corrected post-flash darks from an anneal cycle to determine the average dark level per good pixel,
and mark all hot pixels on shorter timescales with a uniform threshold. This resulted in cleaner and more
accurate images, especially for near ultra-violet and narrow-band images. The CTE-corrected and newly calibrated
science exposures were then combined using AstroDrizzle and an 0.8 pixel fraction at the original pixel scale.

\begin{figure*}
\center{\includegraphics[scale=0.83]{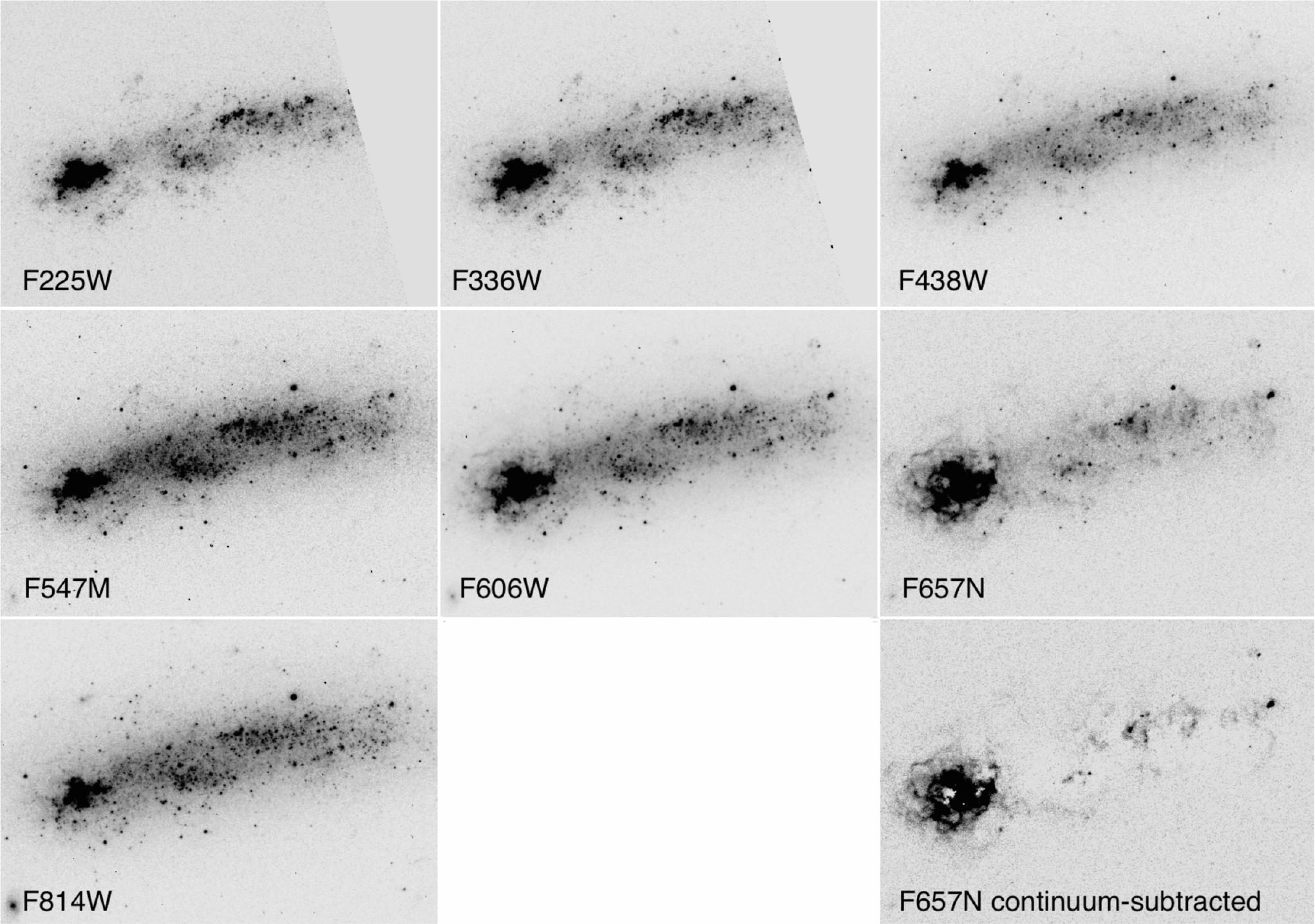}}
\caption{Kiso 5639 images from HST WFC3 observations, with filters indicated.
The lower right-hand figure is the continuum-subtracted H$\alpha$ image.
N is up, E to the left.}\label{filters-new}\end{figure*}

Figure \ref{filters-new} shows the images for the individual filters and for the
continuum-subtracted H$\alpha$ image. The latter was made by scaling the average counts
of the F547M image to the F657N image using the ratio of the PHOTFLAM
values\footnote{\url{http://www.stsci.edu/hst/wfc3/phot_zp_lbn}} for the two filters.
PHOTFLAM is the inverse sensitivity, given in units of erg cm$^{-2}$ \AA$^{-1}$ per
electron count. In practice, this means that the F547M image was divided by 4.78 and
then subtracted from the F657N image to make the continuum-subtracted H$\alpha$ image. We
used the F547M filter for this rather than the F606W filter because the F606W filter
contains the H$\alpha$ line and an equally strong [OIII] line at 5008 $\AA$ according to
the spectrum of this galaxy in the SDSS. The F547M filter lies between these lines. 

\begin{figure}
\center{\includegraphics[scale=0.45]{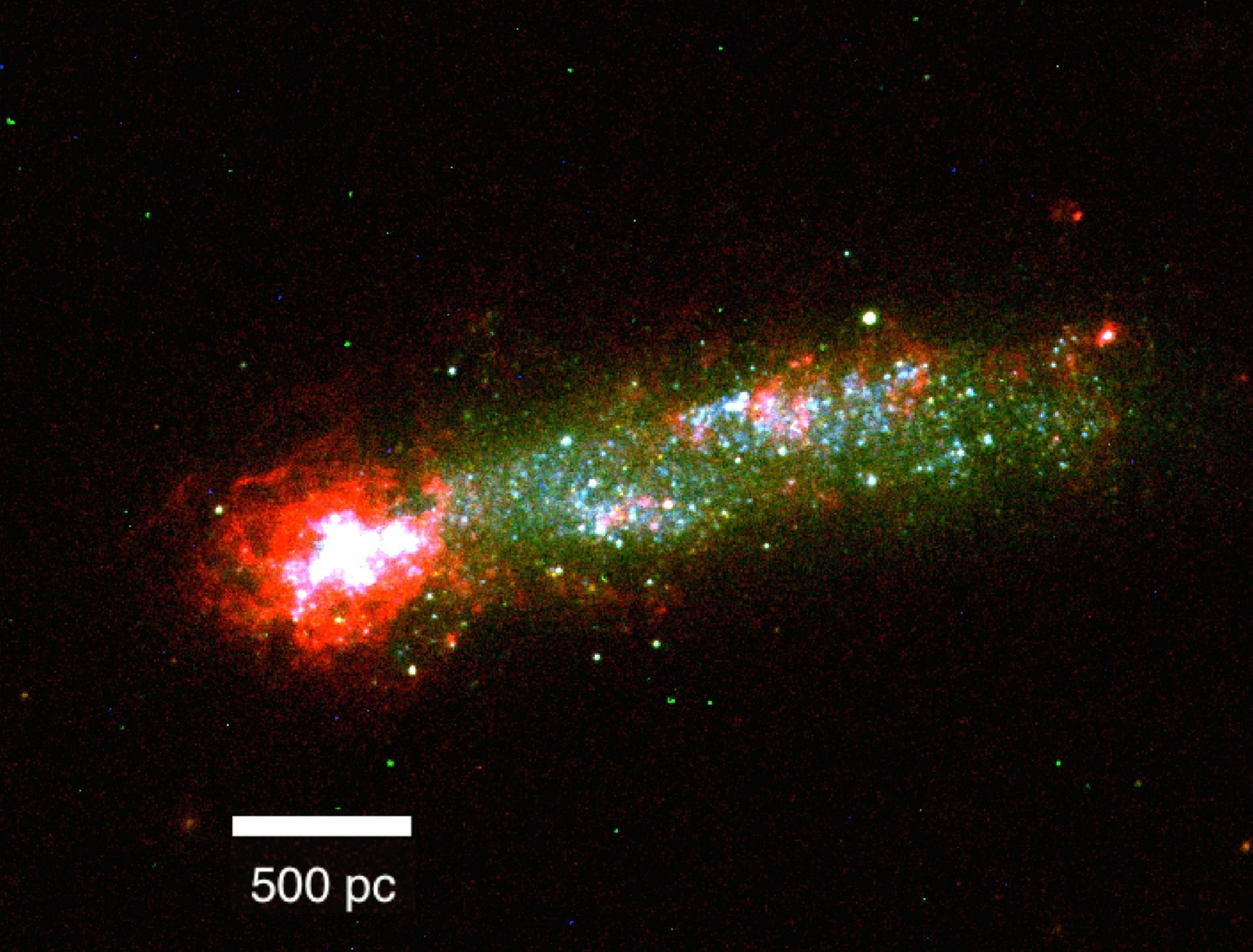}}
\caption{Kiso 5639 color composite image from F657N (in red), F547M (in green), and
F438W (in blue) filters. The line indicates 500 pc,
corresponding to $4.2^{\prime\prime}$. N is up, E to the left.}\label{HaVB}\end{figure}

\begin{figure}
\center{\includegraphics[scale=0.45]{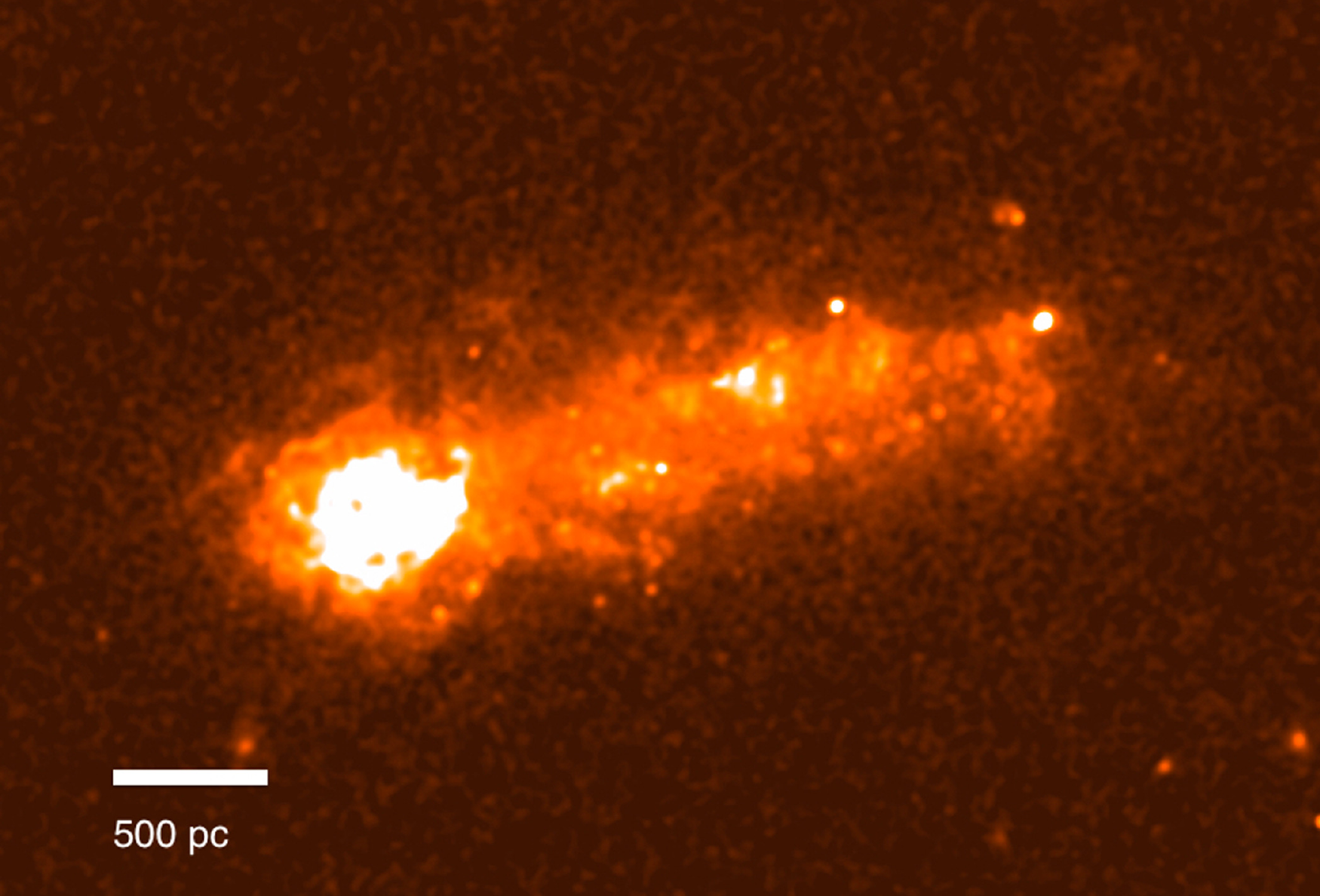}}
\caption{False color image of Kiso 5639 in H$\alpha$ (F657N) (not continuum-subtracted)
on a linear scale, Gaussian smoothed by a factor of 5 in DS9 to enhance the faint peripheral
gas emission. The line indicates 500 pc, corresponding to $4.2^{\prime\prime}$.}\label{hag5}\end{figure}

Color composite figures were made in the Image Reduction and Analysis Facility (IRAF)
using RGB in DS9. Figure \ref{HaVB} shows F438W as blue, F547M as green, and F657N as red
(which is H$\alpha$ including the continuum). The line indicates 500 pc, corresponding to
$4.2^{\prime\prime}$. The periphery of the galaxy is outlined by red wispy protrusions
from H$\alpha$, discussed further in Section \ref{sect:results}. These wisps are shown
better in Figure \ref{hag5}, which is F657N on a linear scale, Gaussian smoothed by a
factor of 5 in DS9 to enhance the faint structures.

\begin{figure}
\center{\includegraphics[scale=0.45]{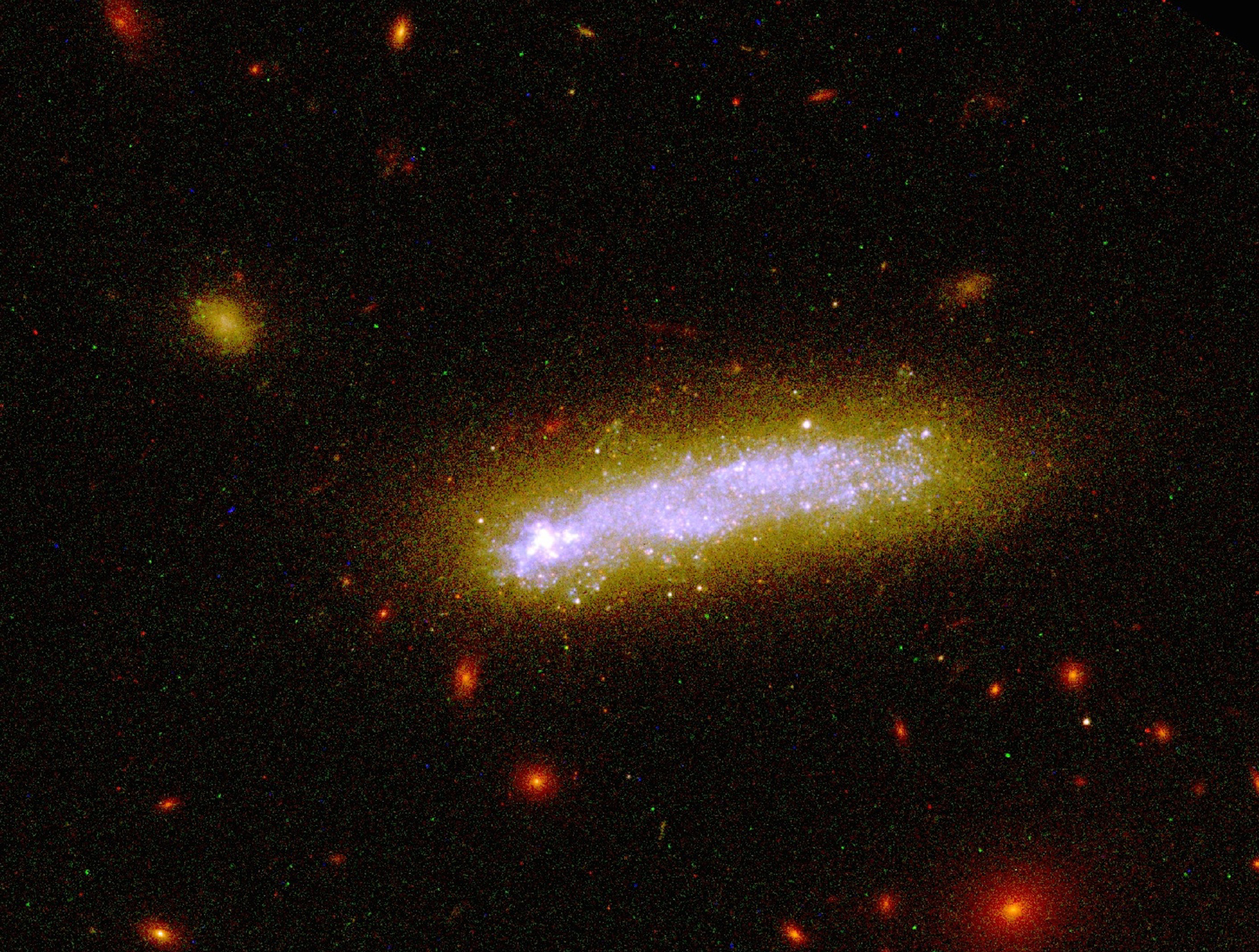}}
\caption{Large field around Kiso 5639 in a color composite image from  F438W (in blue), F547M (in green), and
F814W (in red) filters, shown on a logarithmic scale to enhance the outer boundaries.
Note the small cluster of galaxies surrounding it, and a small galaxy to the northeast.
N is up, E to the left.}\label{largeBVI}\end{figure}

Figure \ref{largeBVI} shows a large field of view with F438W as blue, F547M as green, and
F814W as red. Small background galaxies can be seen, including a larger galaxy with an
unknown redshift to the NE of Kiso 5639. The image is shown on a logarithmic scale to
enhance the extended envelope around the main body of Kiso 5639. Note the extended oval of
yellow color, which is interpreted in Section \ref{round} as an underlying disk. There is
also a small network of intersecting filaments in the north, one-third of the way along the
galaxy from the head to the tail.

\begin{figure}
 \center{\includegraphics[scale=0.4]{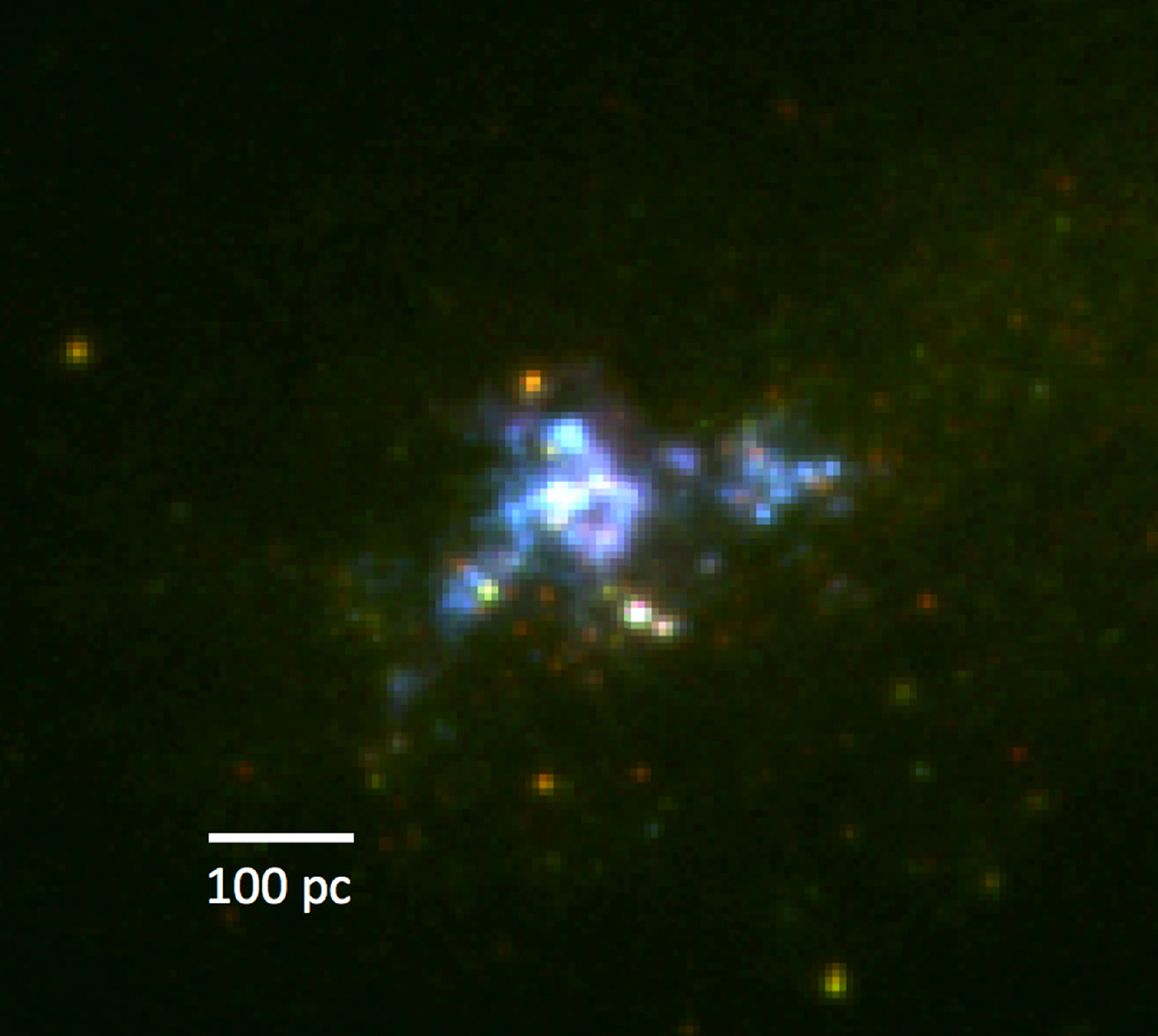}}
\caption{The head region of Kiso 5639 is shown in a color composite image with F336W (blue),
F547M (green), and F814W (red). The white line corresponds to 100 pc = 21.2 px = $0.84^{\prime\prime}$.
The central white cluster is the youngest and most massive in the galaxy.}\label{head}\end{figure}

\begin{figure}
\center{\includegraphics[scale=0.2]{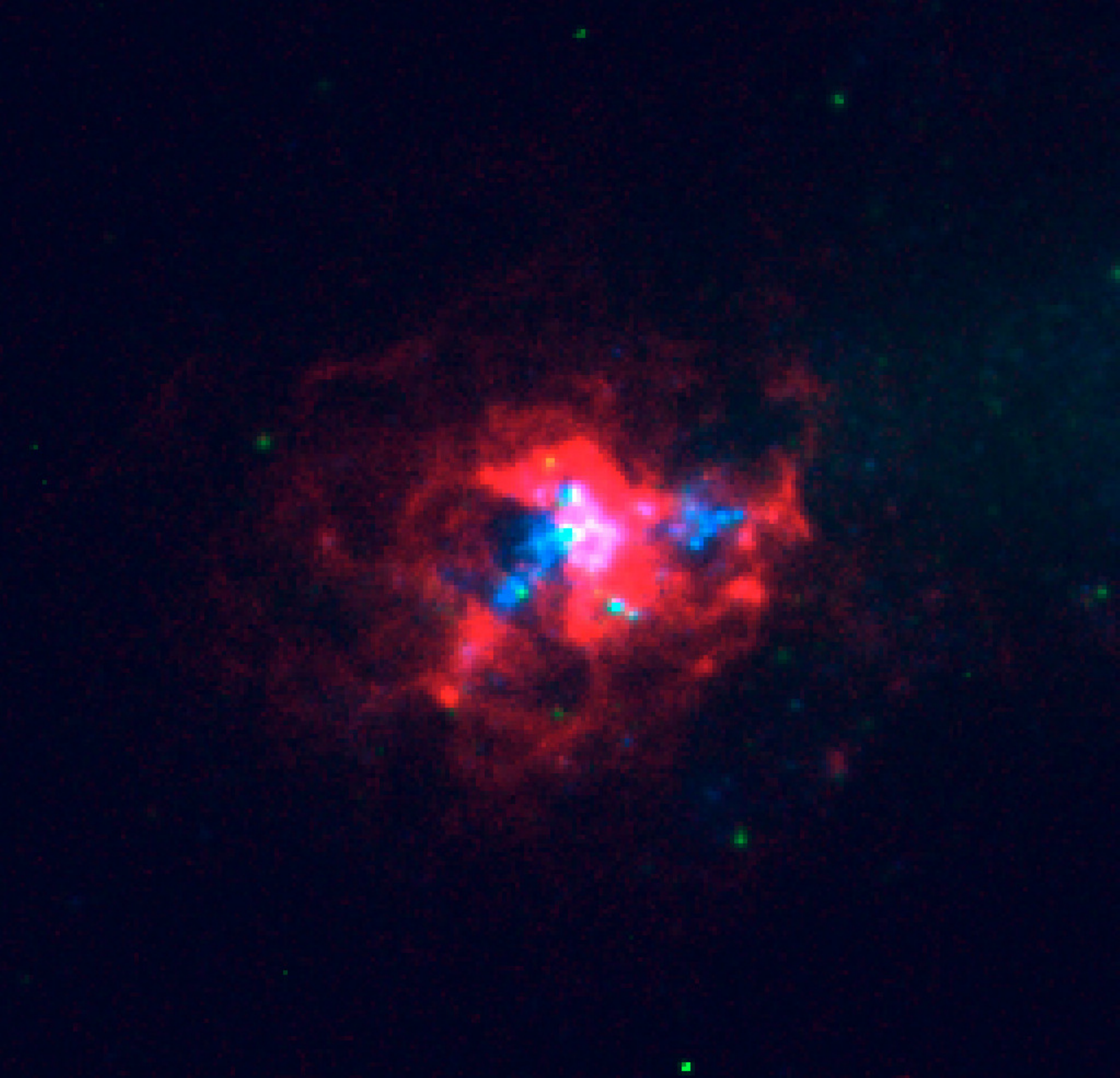}}
\caption{The head region Kiso 5639 is shown in a color composite image with F225W (blue),
F547M (green) and continuum-subtracted H$\alpha$ (red) to highlight the structure of
the emission relative to the young central clusters.}\label{headha}\end{figure}

The head region is shown enlarged in Figure \ref{head}, with F438W in blue, F547M in green, and
F814W in red. The line indicates 100 pc, or $\sim 0.8^{\prime\prime}$. There are many bright
clusters and discrete regions of star formation that are described in the following sections.
In Figure \ref{headha}, the H$\alpha$ continuum-subtracted image is shown in red, with F547M in
green and F225W in blue, to highlight the juxtaposition of young clusters with the H$\alpha$.
There are many cavities in the H\textsc{ii} region (Sect. \ref{holes}).

\section{Analysis}\label{sect:analysis}

\begin{figure}[b!]
\center{\includegraphics[scale=0.42]{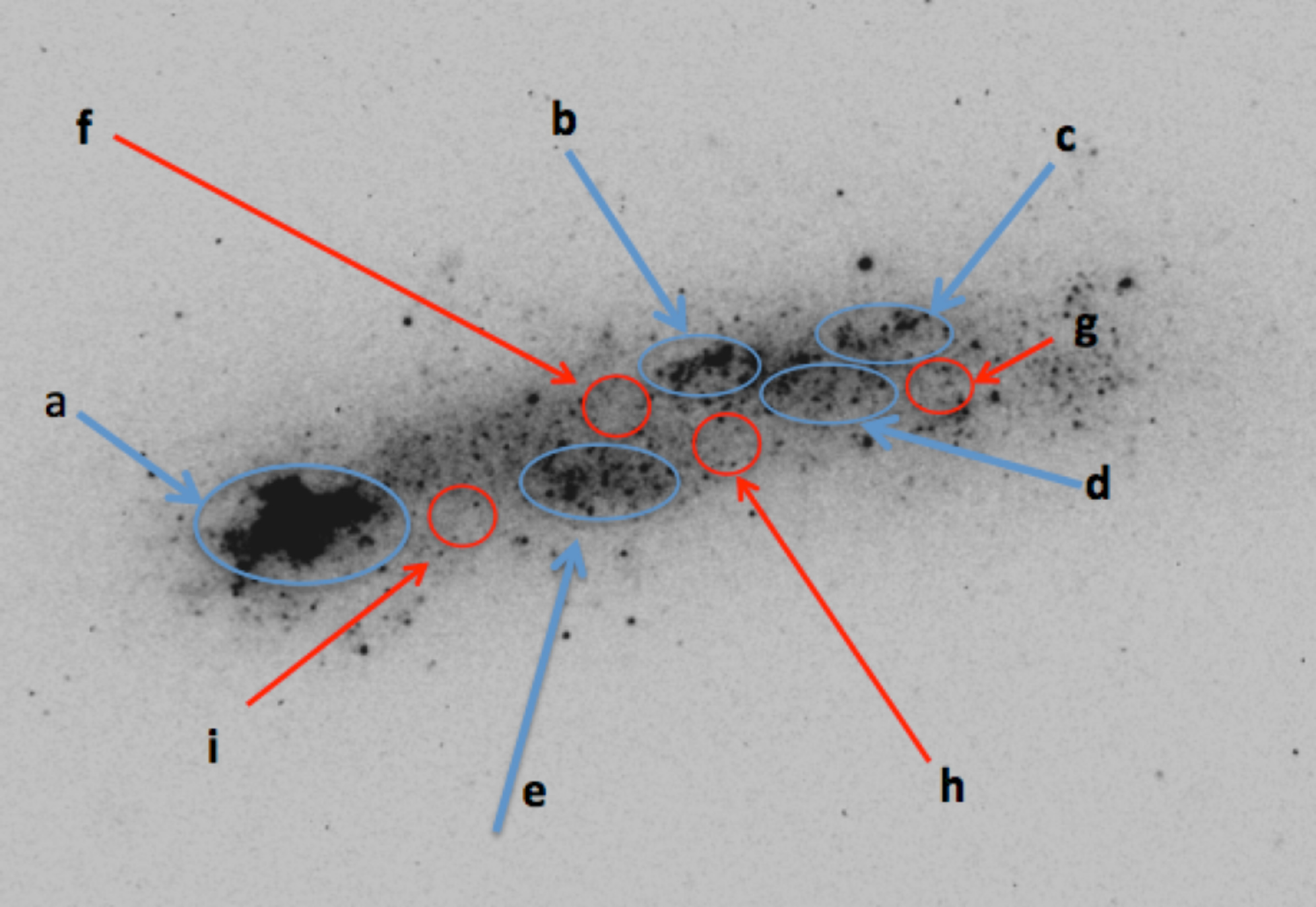}}
\caption{The 5 main regions in Kiso 5639, plus 4 interclump regions, are indicated on the
F438W image.  The head is more massive and younger than the other regions; the interclump
regions are the oldest. Their properties are listed in Table \ref{tab2}.}\label{regions}\end{figure}

Photometric measurements were made of star-forming regions for each image using the task
{\it imstat} in IRAF. Boxes were defined around the 5 most prominent large-scale clumps,
100-200 pc in diameter, based on isophotal contours in the F438W image.  One clump is the
bright part of the head and four are in the tail.  These 5 regions are identified as blue
ellipses in Figure \ref{regions} and labeled $a$ through $e$.  The average interclump background was subtracted from each measurement to get the magnitudes of the individual regions. The interclump surface brightnesses in regions $f$ through $i$ were also determined. Photometric zeropoints from the WFC3
website\footnote{\url{http://www.stsci.edu/hst/wfc3/phot_zp_lbn}} were used to convert
counts to AB mag.

\begin{figure}
\center{\includegraphics[scale=0.48]{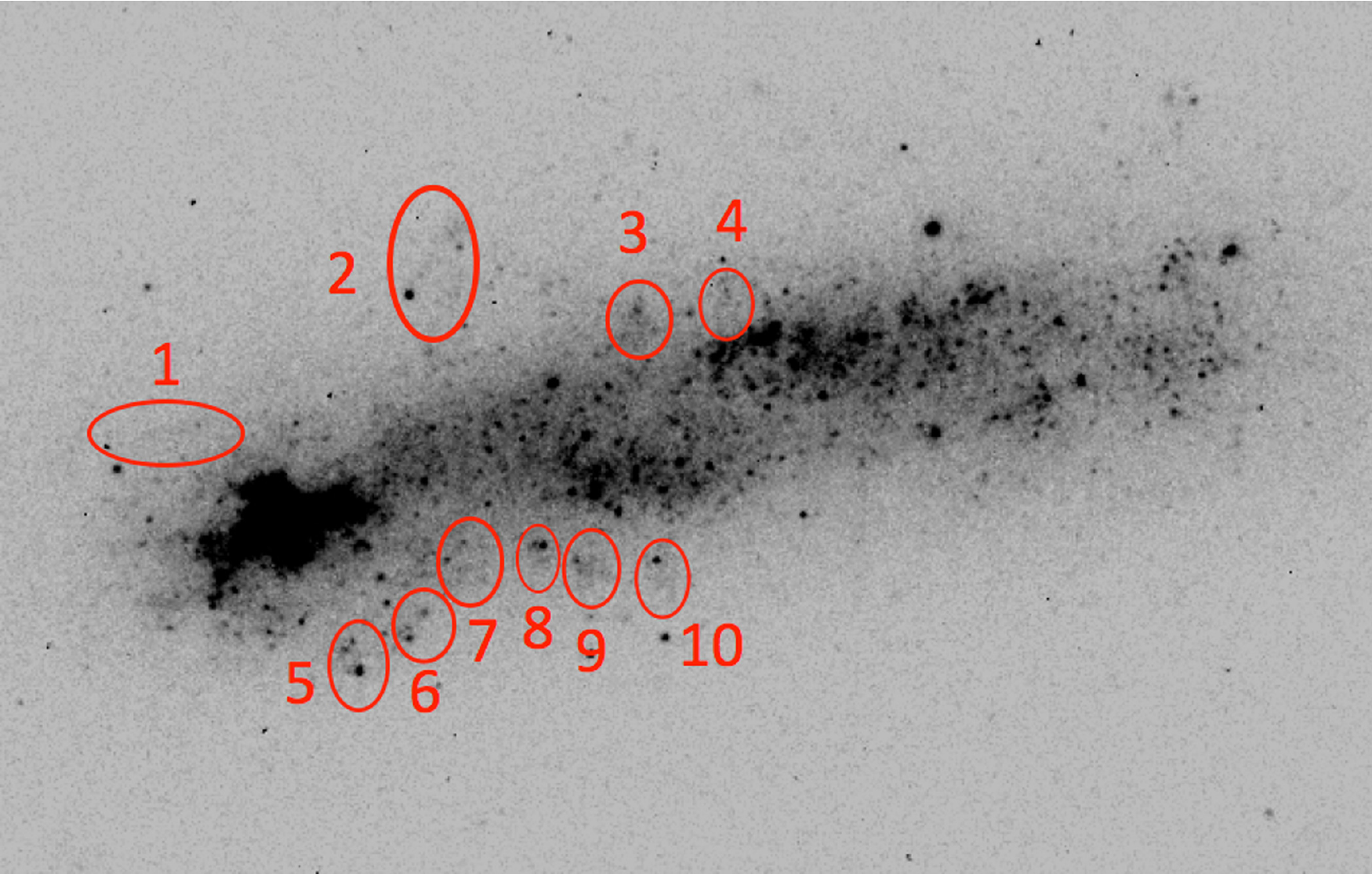}}
\caption{Locations of 10 peripheral filaments or emission regions in Kiso 5639 on the F438W image,
as seen more directly in Figures \ref{HaVB}, \ref{hag5} and \ref{largeBVI}.
The average masses and
ages are in Table 2.}\label{periphery}\end{figure}

In addition, 61 individual
clusters or unresolved complexes 10-20 pc in size were measured in all of these regions, selected by eye from the
F438W images. These clusters are brighter than 25 AB mag in the F438W image. The average interclump background was subtracted from each of these. Ten
filamentary regions extending away from the bright part of the galaxy were measured too,
as outlined in Figure \ref{periphery}, with no background subtraction.

Masses and ages were determined by fitting photometric magnitudes and colors to population
synthesis models from \cite{bruzual} with wavelength-dependent extinction laws from
\cite{calz00} and \cite{leith02}. We followed the procedures in \cite{elm12} and previous
papers, using the integrated population spectra as a function of time as tabulated in
\cite{bruzual}, and the UVIS filters on HST. Throughputs for the UVIS filters were from the
WFC3
website\footnote{\url{http://www.stsci.edu/hst/wfc3/ins_performance/throughputs/Throughput_Tables}}.
The Chabrier initial mass function (IMF) and a metallicity of 0.2 solar were assumed. For the
individual clusters, color models were generated, with the age of the cluster corresponding to
the beginning of star formation, and an exponential decay going forward. Models for the large
star-forming regions and the interclump regions were made assuming a constant SFR back to the
fitted age.

Many models were run for each region, covering all combinations of visual extinction from
0 to 7.5 mag in steps of 0.25 mag, logarithmic age (in years) in 50 equal steps from 5.1
to 10, and the decay times of 1, 3, 10, 30 and 100 times $10^7$ years. Each combination
produced theoretical colors that were compared to the observed colors using the quadratic
sum of differences, $\delta^2$.  This $\delta^2$ is different from $\chi^2$ by not having
the rms noise in the denominator of each difference. The rms noise from observations is
much less than the rms variation from model to model. The region masses were determined
from the models for each case also, scaling the theoretical magnitude in F438W to the
observed magnitude and using the masses tabulated for each population and age by
\cite{bruzual}. We then picked as reasonable solutions those with the lowest $\delta^2$,
and among those took the averages of the masses, ages, extinctions and decay times using
a weighting function of $\exp(-0.5\delta^2)$.

The metallicity measurements of Kiso 5639 by \cite{jorge13} are 0.1 solar for the head and 0.2
solar for the tail, so the tabulated Bruzual models with a metallicity of 0.2 solar (Z=0.004)
are reasonable. To determine how important metallicity was, solutions for the individual
clusters were also found for solar and 0.4 solar metallicities. For the 21 head clusters, the
log-masses were on average 4.2\% and 1.5\% smaller than our measurements for solar and 0.4
solar metallicity, their log-ages were 3.9\% and 3.4\% larger, their extinctions were 16.6\%
and 9.2\% smaller, and their SFR decay log-times were less than 1\% larger, respectively.
These changes are small enough to ignore metallicity effects on the photometric analysis.

Star formation rates (SFRs) were determined from the continuum-subtracted H$\alpha$ image
for the large-scale regions. Taking into account the distance and the filter width (121
$\AA$), the luminosity is given by $L(H\alpha)=2.19\times 10^{46} \times
10^{-0.4m_\alpha}$ erg s$^{-1}$ for apparent AB H$\alpha$ magnitude, $m_\alpha$.
\footnote{The coefficient is $4\pi D^2 \Delta \lambda c/\lambda^2
10^{-0.4\times48.6}$ for distance D, filter width $\Delta \lambda$, speed of light $c$
and wavelength $\lambda$.} The SFRs were determined from $SFR=7.9\times 10^{-42}
L(H\alpha)$ $M_\odot$ yr$^{-1}$ using the calibration from \cite{ken}.

\section{Results}
\label{sect:results}

\subsection{Masses, Ages, and Star formation rates}
\label{masagesfr}

Table \ref{tab2} lists the mass and age solutions from the photometry for the main
large-scale star-forming clumps in the head and tail (regions $a-e$ in Figure
\ref{regions}), for the averages of all the individual clusters in the head and tail
regions, for the interclump regions ($f-i$), and for the averages of the filamentary
protrusions on the periphery of the galaxy in the head (ellipse 1 in Figure
\ref{periphery}) and in the tail (ellipses 2 - 9). The SFRs are the ratios of the
photometric masses to the ages, the surface densities are the ratios of the masses to the
areas used for the photometry, and the H$\alpha$ luminosities and H$\alpha$ SFRs are from
the continuum-subtracted H$\alpha$ filter, evaluated only for the large regions.

\begin{deluxetable*}{lcccccc}
\tabletypesize{\scriptsize}\tablecolumns{7} \tablewidth{0pt} \tablecaption{Masses, Ages, and Star Formation Rates}
\tablehead{
\colhead{Region}&
\colhead{log Mass} &
\colhead{log Age} &
\colhead{log SFR}&
\colhead{Surface density}&
\colhead{L(H$\alpha$)}&
\colhead {log SFR(H$\alpha$)}\\
\colhead{}&
\colhead{(M$_{\odot}$)} &
\colhead{(yr)} &
\colhead{(M$_{\odot}$ yr$^{-1}$)}&
\colhead{(M$_{\odot}$ pc$^{-2}$)}&
\colhead{$\times 10^{37}$ erg s$^{-1}$}&
\colhead{(M$_{\odot}$ yr$^{-1}$)}}
\startdata
Main clumps\tablenotemark{a,b}:	&			&			&			&		&		&		\\
a (head)	&	6.2	$\pm 0.12$	&	5.7	$\pm 0.39$	&	 0.43	$\pm 0.41$	&	7.8	&	$880\pm12$	&	-1.2	\\
b (tail)	&	5.2	$\pm 0.14$	&	6.6	$\pm 0.23$	&	-1.4	$\pm 0.27$	&	4.3	&	$13\pm0.1$	&	-3.0	\\
c (tail)	&	5.0	$\pm 0.10 $	&	5.8	$\pm 0.64$	&	-0.7	$\pm 0.64$	&	2.9	&	$14\pm0.3$	&	-3.0	\\
d (tail)	&	5.2	$\pm 0.10 $	&	6.7	$\pm 0.05$	&	-1.5	$\pm 0.02$	&	3.1	&	$8.6\pm0.2$	&	-3.2	\\
e (tail)	&	5.6	$\pm 0.11$	&	7.5	$\pm 0.63$	&	-1.9	$\pm 0.64$	&	3.0	&	$18\pm0.3$	&	-2.8	\\
Clusters\tablenotemark{c}:	&			&			&			&		&		&		\\
a (head, 21)	&	4.2	$\pm 0.28$	&	5.9	$\pm 0.88$	&	-1.7	$\pm 0.92$	&	66$\pm 25$	&		&		\\
b (tail, 10)	&	3.7	$\pm 0.46$	&	7.0	$\pm 0.93$	&	-3.3	$\pm 1.04$	&	23$\pm 18$	&		&		\\
c (tail, 9)  	&	3.5	$\pm 0.32$	&	6.8	$\pm 0.74$	&	-3.4	$\pm 0.80$	&	15$\pm 5.7$	&		&		\\
d (tail, 9)	    &	3.5	$\pm 0.36$	&	6.7	$\pm 0.78$	&	-3.3	$\pm 0.86$	&	16$\pm 7.4$	&		&		\\
e (tail, 12)	&	3.5	$\pm 0.48$	&	6.7	$\pm 0.95$	&	-3.2	$\pm 1.07$	&	43$\pm 77$	&		&		\\
Interclumps\tablenotemark{d}:	&			&			&			&		&		&		\\
f	&	5.2	$\pm 0.07$	&	8.9	$\pm 0.51$	&	-3.7	$\pm 0.52$	&	26.3	&		&		\\
g 	&	4.9	$\pm 0.03 $	&	9.0	$\pm 0.39$	&	-4.0	$\pm 0.39$	&	18.2	&		&		\\
h	&	5.3	$\pm 0.09$	&	9.1	$\pm 0.55$	&	-3.8	$\pm 0.56$	&	29.8	&		&		\\
i 	&	5.6	$\pm 0.03$	&	9.1	$\pm 0.36$	&	-3.5	$\pm 0.36$	&	28.3	&		&		\\
Periphery\tablenotemark{e}:	&			&			&			&		&		&		\\
1 (head) 	&	4.6	$\pm 0.12$	&	5.8	$\pm 0.65$	&	-1.2	$\pm 0.66$	&	8.7	&		&		\\
2-10 (tail)	&	4.5	$\pm 0.42$	&	7.6	$\pm 0.89$	&	-3.1	$\pm 0.98$	&	5.8$\pm2.7$	&		&		
\label{tab2}
\enddata
\tablenotetext{a}{SED fits for the head removed line emission from the F606W flux.  A
constant star formation rate back to the indicated age was assumed for these large-scale
regions.}
\tablenotetext{b}{The masses and ages for the clump regions correspond to
rectangles with areas of 8085, 1512, 1566, 2160, and 5252 pixels for a, b, c, d, and e,
respectively. The SED fits are an average of the fits with the lowest rms deviations from
the observed colors considering exponentially decaying star formation rates starting at
the indicated age in the past and decaying with timescales of $10^7$, $3\times10^7$,
$10^8$, $3\times10^8$ and $10^9$ years.}
\tablenotetext{c}{The numbers in parentheses are the number of clusters (or complexes) measured in each region.}
\tablenotetext{d}{The masses and ages for the
interclump regions correspond to rectangles with areas of 255, 204, 270, and 609 pixels
for f, g, h, and i, respectively. A constant star formation rate back to the indicated
age was assumed.}
\tablenotetext{e}{A constant star formation rate back to the indicated
age was assumed.}
\end{deluxetable*}

Uncertainties for the solutions were determined from the three sources of uncertainty added in
quadrature. The first is  the dispersion in the best fit log-values for each cluster in a
region. The second is the variation in the model results (in logarithms) among all of the
individual solutions for each cluster that were combined with the $\exp(-0.5\delta^2)$
weighting factor to give the best fit (see above). The third is the measurement error from the
standard deviation of the image counts relative to the average counts in each region, summed in
quadrature for each filter.

Extinction results are not given in the table but they average about 0.6-0.8 mag visual for
all of the clusters in the head and tail. The average decay rate in the best fit cluster
solutions is $\sim10^8$ years, which is typically longer than the cluster age, meaning that
the preferred SFRs are fairly constant.

\begin{figure}
\center{\includegraphics[scale=0.47]{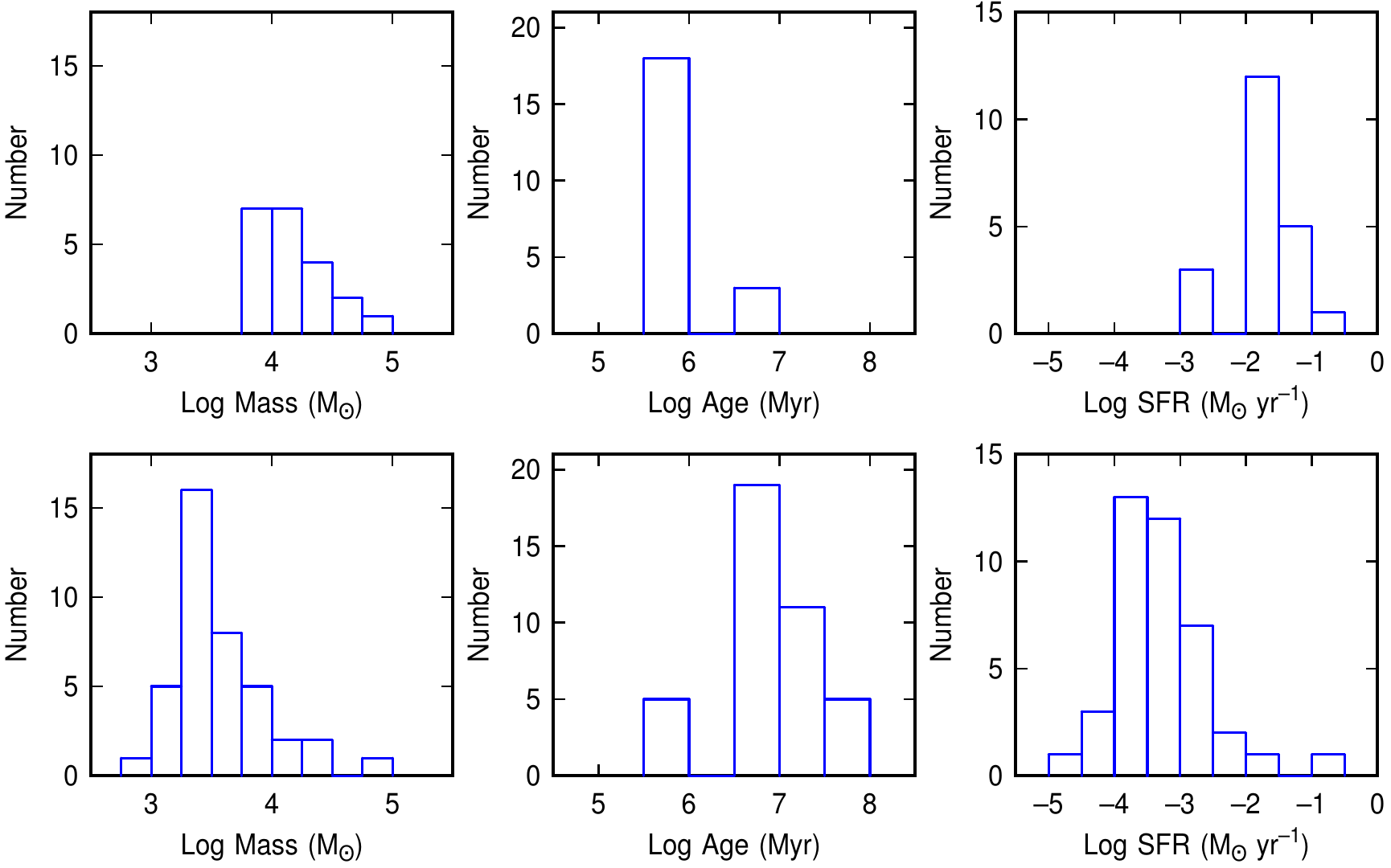}}
\caption{Mass, age, and star formation rate histograms for the star-forming clusters for the head (top) and tail (bottom) regions.}\label{tadpolhis2}\end{figure}

Figure \ref{tadpolhis2} shows histograms for mass, age, and SFR for the 21 clusters in the
head (top panels) and the 40 clusters in the 4 main clumps in the tail (regions $b-e$). The
individual clusters in the head have average masses of $1.6\times10^4 M_{\odot}$, which is
3-6 times larger than those in the tail. The ages of the head clusters average $\sim
8\times10^5$ yr, or about 7-12 times younger than in the tail.

\begin{figure}
\center{\includegraphics[scale=0.6]{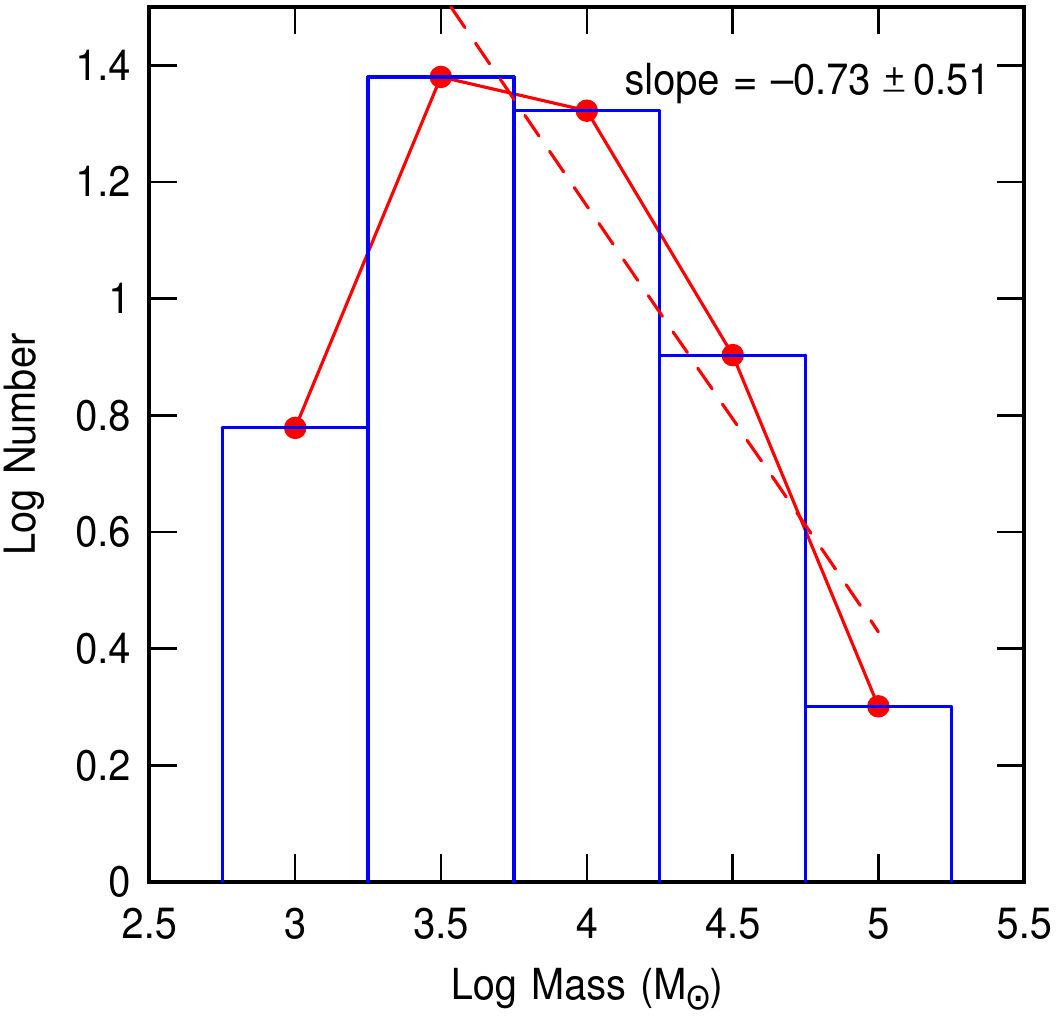}}
\caption{Mass histogram and power law fit for  the 61 measured star-forming
clusters in Kiso 5639.}\label{masshistall}\end{figure}

A histogram of mass for all 61 measured clusters is shown in Figure \ref{masshistall} with
a power law fit at $\log M\ge3.5$.  The detection limit is $\sim3\times10^3\;M_\odot$. The
slope of the mass function is $-0.73\pm0.51$ on this log-log plot, which corresponds to a
slope of $-1.73\pm0.51$ in linear intervals of mass, i.e., $dN/dM\propto M^{-1.73\pm0.51}$.
The error bars are the 90\% confidence interval for a student t test with 4 degrees of
freedom. The correlation coefficient is $r=0.99$.  This slope is typical for cluster mass
functions \citep{larsen09}.

\begin{figure}
\center{\includegraphics[scale=0.6]{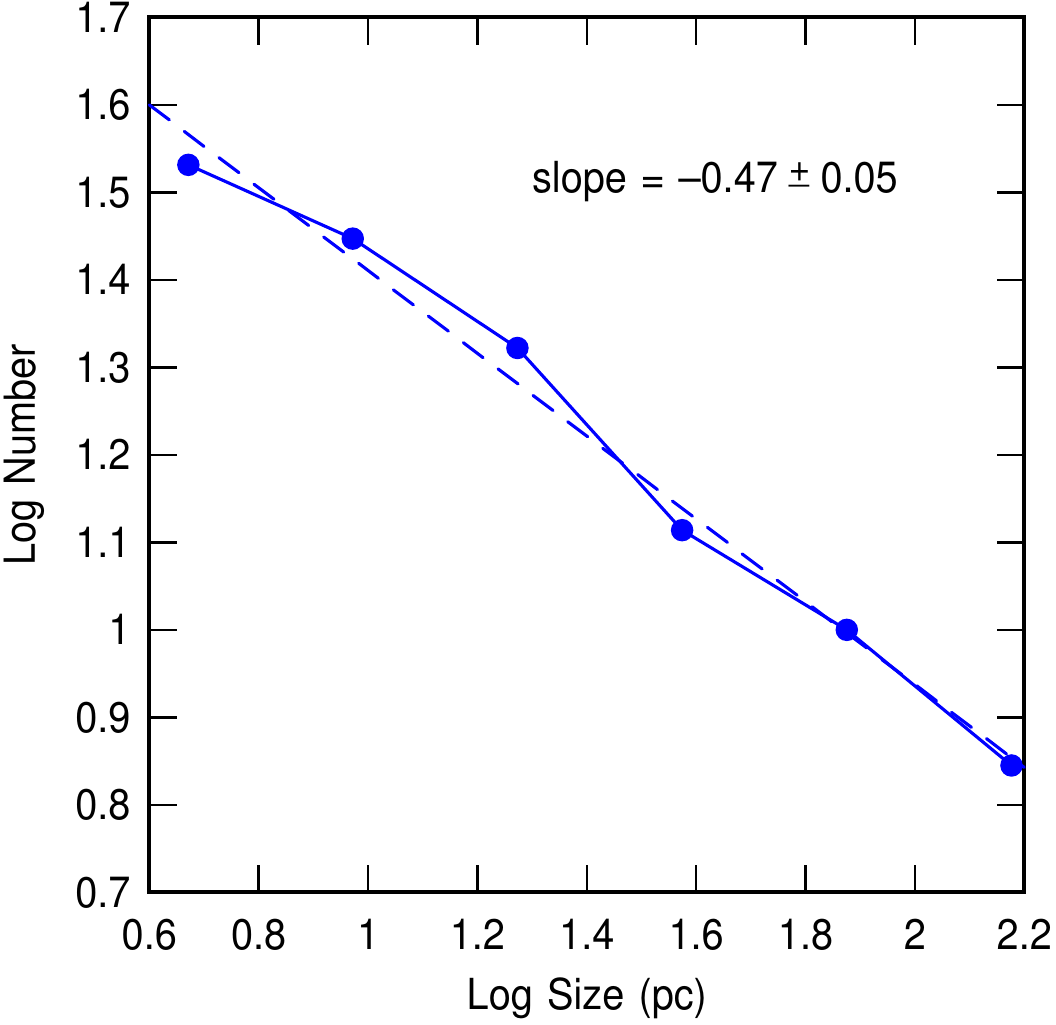}}
\caption{Histogram of number versus size of the star-forming clumps, as measured from
SExtractor on images Gauss-blurred by 1, 2, 4, 8, 16, 32.}\label{logNlogS}\end{figure}

Star-forming regions in galaxies generally follow a hierarchical size distribution
\citep{elmegreen10}. To check the size distribution of the star-forming regions in Kiso
5639, we used the IRAF task {\it gauss} to Gaussian blur the F336W image by five successive
factors of two. Then the program SExtractor was used to identify and count sources in each
of these Gauss-blurred images. The result is a count of regions larger than the blurring
size. A plot of the log-count versus the minimum log-size is given in Figure
\ref{logNlogS}. The power law distribution from 4 pc to 150 pc follows from the
hierarchical structure; here the slope is $-0.47\pm0.05$. For 12 spiral and irregular
galaxies in \cite{elmLEG}, a similar procedure gave slopes of $-0.76$ to $-1.86$ using the
LEGUS Legacy ExtraGalactic UV Survey \citep{calz15}. A shallow slope, like that found here,
means that star formation is spread out somewhat uniformly. This is reasonable from Figure
\ref{largeBVI}, where the main star-forming regions are spread broadly across the galaxy.

\subsection{Tadpole Intrinsic Shape}
\label{round}

\begin{figure}
\center{\includegraphics[scale=0.45]{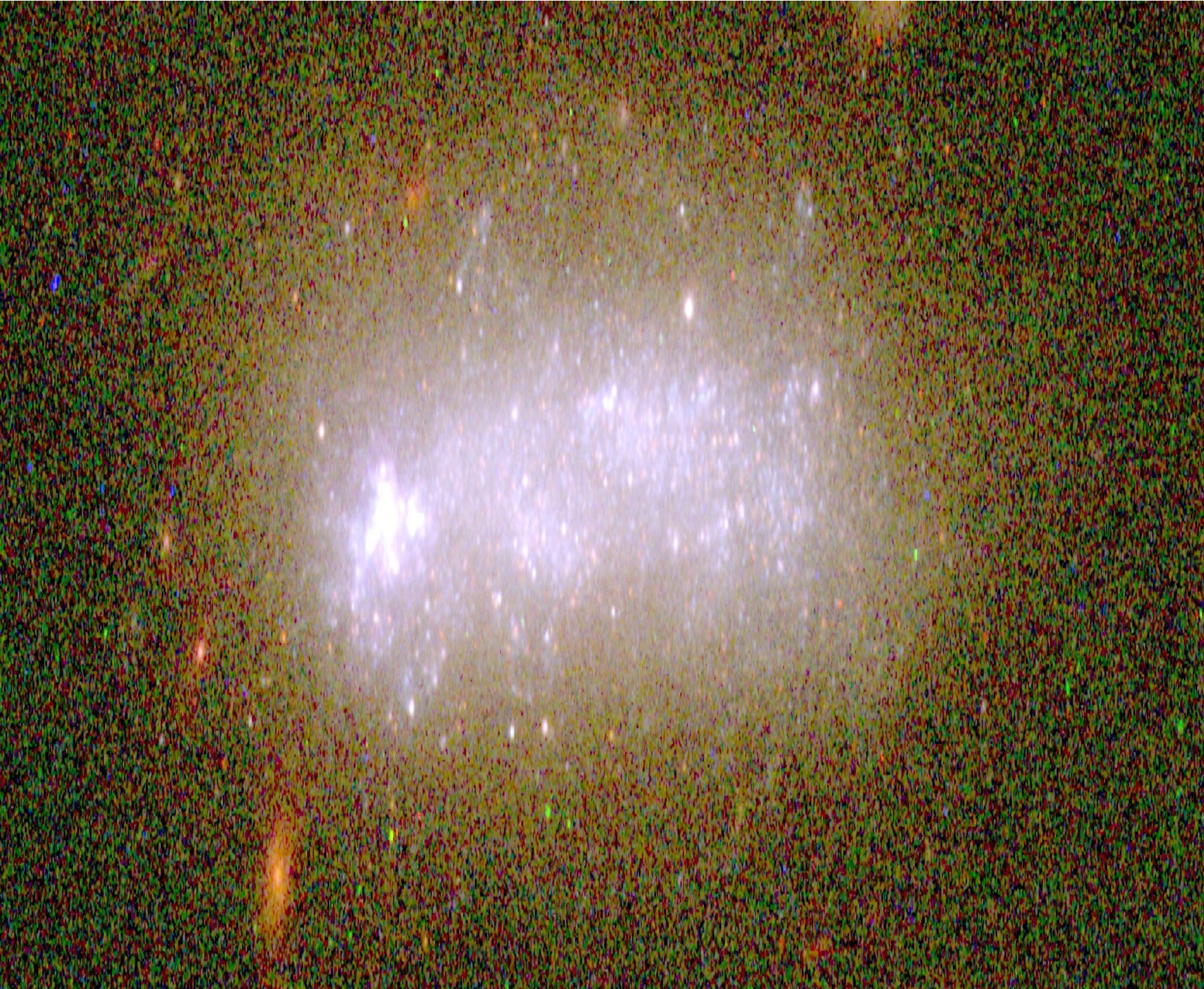}}
\caption{Deprojected image of F438W (blue), F547M (green) and F814W (red),
assuming that the galaxy is round and using a rotation of $-9.3^{\circ}$ and an inclination of 68$^{\circ}$ as determined from the outer F814W isophotes.}\label{deproject}\end{figure}

The three-dimensional structure of Kiso 5639 is not easily determined from its projected
shape, but the optical spectrum implies rotation \citep{jorge13}.  The F814W image also
shows what may be a faint underlying disk, especially in the logarithmic stretch in Figure
\ref{largeBVI}. Based on the assumption that the main disk is intrinsically round, we
measured the outer isophotes of the major and minor axes of the F814W image to determine
the orientation and inclination. The images were then rotated in IRAF by -9.3$^{\circ}$
using the task {\it rotate} and stretched by a factor of 2.67 (corresponding to an
inclination of 68$^{\circ}$ for a thin disk) using the task {\it magnify}. Figure
\ref{deproject} shows the deprojected color composite image for the galaxy using F438W
(blue), F547M (green), and F81W (red) images. The head region is round in the inclined view
and elongated in the deprojected view. The projected roundness suggests that the head is
more three-dimensional like a sphere than two-dimensional like a circle in the plane. Disk
thickness elsewhere will cause another distortion, with one side of the minor axis showing
slightly more stars from the back side of the disk and the other side showing slightly more
from the front side. If we correct the apparent inclination, $i_{\rm appar}$, for an
intrinsic ratio of thickness to diameter equal to 0.2 using the equation $\cos i_{\rm
true}^2=(\cos i_{\rm appar}^2-0.2^2)/(1-0.2^2)$ from \cite{sanchezjanssen10}, then the true
inclination is $i_{\rm true}=84^\circ$.

The assumption of a circular disk for the whole galaxy is reasonable considering axial
ratios for one-clump galaxies in the Kiso sample of UV bright galaxies \citep{kiso-all}. If
tadpoles are highly inclined disk galaxies, there should be face-on versions of tadpoles
that have one dominant star-forming clump in a round disk. Kiso galaxies labeled as I,c and
I,g are galaxies identified as having one or several main clumps. There are 159 of these
galaxies in the Sloan Digital Sky Survey (SDSS), which we examined to select one-clump
galaxies. Some galaxies in these lists were clearly spiral or interacting, and were
excluded. Of the remainder, 48 had one main clump, and 44 had several main clumps; some of
the one-clump dominant galaxies were in the I,g category.

\begin{figure}[]
\center{\includegraphics[scale=0.7]{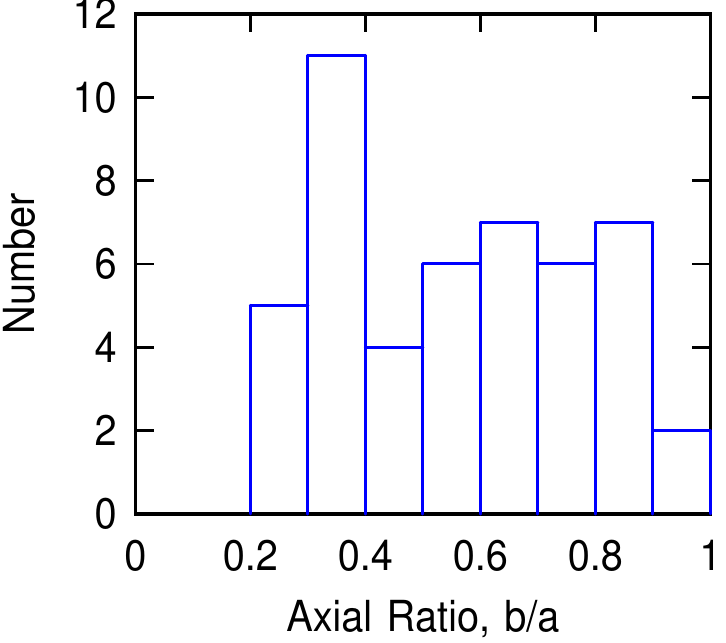}}
\caption{Axial ratios of galaxies from the Kiso ultraviolet survey with one main star-forming clump, measured from
SDSS images. }\label{clumphist}\end{figure}

A histogram of the isophotal axial ratios for the 48 one-clump galaxies is shown in
Figure \ref{clumphist}. The distribution is fairly flat over most of the range, except
for a spike at an axial ratio of 0.35. This flatness means that most one-clump galaxies
are approximately circular disks seen in random projections with tadpoles representing
the edge-on cases \citep[compare to similar distributions for clumpy galaxies
in][]{elmegreen05a,elmegreen05b}. The spike suggests that a small fraction of one-clump
galaxies may be intrinsically elongated. For example, the galaxy I Zw 18 is an apparent
tadpole composed of one large clump of star formation in a main body with a smaller one
to the south in the same old-stellar envelope, and a third clump to the north, somewhat
detached; it appears to be intrinsically elongated.  The drop-off in the distribution of
axial ratios below 0.2 in the figure is presumably the ratio of disk thickness to
diameter. Small galaxies like this tend to be relatively thick
\citep{elmegreen05b,sanchezjanssen10,eh15}.

\subsection{Star-forming Head}

\subsubsection{Star Formation and Ionization Rates}

The H$\alpha$ luminosity of the central $3^{\prime\prime}$ diameter region of the head
(360 pc) is 8.0x10$^{39}$ erg s$^{-1}$ according to the SDSS spectrum \citep{elm12}. 
For the HST observations, the H$\alpha$ luminosity of a core rectangular region centered
on the head is $8.8\pm0.16\times10^{39}$ erg s$^{-1}$. This H$\alpha$ luminosity is
determined from the continuum-subtracted F657N H$\alpha$ image, which comes from the
H$\alpha$ filter width, $121\AA$, multiplied by the difference between the F657N
intensity and the F547M intensity used for the continuum, and then multiplied by $4\pi
D^2c/\lambda^2$ (see footnote 6 above).  For reference, the continuum-subtracted
H$\alpha$ apparent AB magnitude of the head is $15.99$. The size of this region is
effectively $3.6^{\prime\prime}$, calculated as the square root of the number of pixels
in the rectangle multiplied by the pixel size. Assuming the conversion in equation (2) of
\cite{ken}, the head luminosity from HST H$\alpha$ corresponds to a SFR of
$\sim0.070\;M_{\odot}$ yr$^{-1}$, which gives the log value of $-1.2$ in Table 2.

The SFR determined from the ratio of the head region's photometric mass to its
photometric age, $2.7\;M_\odot$ yr$^{-1}$, is much larger than the H$\alpha$ rate. One
reason for this may be that the photometric ages are too young compared to the overall
duration of star formation in the region. This could be the result of a dominance by the
youngest massive stars in the ultraviolet bands, or from patchy dust that preferentially
obscures older stars. There could also be a component of H$\alpha$ that is obscured or
too extended to see in our data. Extinction has an important effect on the inferred
photometric SFR. Higher extinction requires more stellar mass to give the same
brightness, and a younger age to give the same color, and the combined effect pushes up
the SFR. Varying the extinction in our models from 0 to 1 magnitude in the visible band
increases the total photometric SFR for the 21 clusters in the head from $0.13\;M_\odot$
yr$^{-1}$ to $1.2\;M_\odot$ yr$^{-1}$.  The average cluster value in Table 2, $10^{-1.7}$
per cluster, corresponds to a total for all 21 clusters of $0.4\;M_\odot$ yr$^{-1}$, in
the middle of this range and consistent with the average extinction that we find for
these clusters, which is 0.65 mag.  We do not derive the SFR based solely on the FUV flux
because that method needs to assume a star formation history, such as a constant rate. In
what follows, we use the SFR from the H$\alpha$.

For typical recombination in a photoionized region, the Lyman continuum photon ionization
rate, $S$, in photons s$^{-1}$, is proportional to the H$\alpha$ luminosity in erg
s$^{-1}$ according to the conversion $S=7.3\times10^{11} L(H\alpha)$, again using
equation (2) in \cite{ken}. This implies the ionization rate is
$\sim6.4\times10^{51}$ photons s$^{-1}$. For an ionization rate from an O9 star of
$3\times10^{48}$ s$^{-1}$ \citep{vacca96}, the head of Kiso 5639 contains $\sim2100$
O9 stars.

The summed mass of the 21 measured clusters in the head region is
$4.0\times10^5\;M_\odot$. Considering a Chabrier IMF in the 0.2 solar metallicity models
by \cite{bruzual}, such a mass of young stars at 1 Myr age would be expected to have an
ionization rate of $1.9\times10^{51}$ s$^{-1}$; at 3 and 10 Myr, that would be $1.5$ and
$0.4$ times $10^{51}$ s$^{-1}$, respectively. The rate for the 1 Myr age is about 30\% of what we
observe. Alternatively, we can integrate an IMF that is flat below $0.5\;M_\odot$ and a
power law with the Salpeter slope of $-2.35$ above $0.5\;M_\odot$ to $120\;M_\odot$, and
normalize it to the same total cluster mass. This gives $\sim900$ stars more massive than
$25\;M_\odot$, the mass of an O9 star \citep{vacca96}; that is $\sim40$\% of the
number obtained from the total ionization rate. These fractions imply that 30\%-40\% of
the O stars in the head of Kiso 5639 are in clusters; the rest are presumably between the
clusters.

\subsubsection{Ionization and Confinement of the Head H\textsc{ii} Region}
\label{sect:confine}

If the full diameter of the head, $7^{\prime\prime}\sim830$ pc, is the diameter of a
Str\"omgren sphere formed by ionization at the inferred rate of $6.4\times10^{51}$
photons s$^{-1}$, then the rms electron density is $n_{\rm e}\sim1.7$ cm$^{-3}$, assuming
a recombination coefficient $\alpha^{(2)}=2.6\times10^{-13}$ cm$^3$ s$^{-1}$. At a
temperature $T=10^4$ K, the corresponding pressure would be $P_{\rm
H\textsc{ii}}=2.1n_ek_{\rm B}T=3.5\times10^4k_{\rm B}$ for Boltzmann's constant $k_{\rm
B}$.

We are interested in whether this pressure is too large for the galaxy to retain
the gas. If the pressure exceeds the self-gravitational pressure from the weight of the
disk, then the H\textsc{ii} region will expand. For the observed H$\alpha$ line
half-width of $\sim23$ km s$^{-1}$ \citep{jorge13}, the expansion time over the radius of
415 pc is only $\sim20$ Myr. A recent WIYN telescope spectrum taken by one of us (JSG)
suggests that the H$\alpha$ line $\sigma$ is $\sim30$ km s$^{-1}$, which would make the
expansion time 14 Myr. These high linewidths for the ionized gas also suggest that the
H\textsc{ii} region pressure is larger than $\sim3.5\times10^4k_{\rm B}$ from the
thermal motions alone. Gas with a density of $\rho=1.7 m_{\rm H}\mu$ cm$^{-3}$ for
hydrogen mass $m_{\rm H}$, mean atomic weight $\mu=1.36$, and turbulent speed
$\sigma\sim30$ km s$^{-1}$ has a pressure $\rho \sigma^2\sim2.5\times10^5\;k_{\rm B}$.

The self-gravitational pressure in a disk is approximately $P_{\rm
dyn}=(\pi/2)G\Sigma_{\rm tot}^2$ for total mass surface density $\Sigma_{\rm tot}$, which
consists of stars, gas, and some dark matter \citep{elmegreen89}. The average surface
density of old stars in the interclump regions is $\sim25\pm5\;M_\odot$ pc$^{-2}$, as
obtained from SED fitting (Table \ref{tab2}). This is about the same as the surface
density obtained from the galaxy dynamical mass, $1.5\times10^8\;M_\odot$ (Section
\ref{sect:reduc}) spread out over the galaxy diameter of 2.7 kpc, assuming the galaxy is
a circular disk. That result is $26\;M_\odot$ pc$^{-2}$, although it probably includes an
amount of dark matter comparable to the stellar mass. The SED mass of the head
without the background interclump emission subtracted (as it is in Table 1), is
$3.1\times10^6\;M_\odot$ inside the area measured (8085 square pixels or $1.79\times10^5$
pc$^2$). This gives a total stellar surface density of $17.3\;M_\odot$ pc$^{-2}$ in the
head region. The dynamical mass in the 130 pc diameter core of the head measured from the
H$\alpha$ velocity gradient \citep{jorge13} is $1.7\times10^7\;M_\odot$.  This suggests a
surface density of $1.3\times10^3\;M_\odot$ pc$^{-2}$, which is much larger than the
photometric surface density. The difference could be the result of a component in the
H$\alpha$ velocity gradient that is caused by pressure forces, rather than gravity, or it
could indicate a dense dark mass in this region, perhaps from a molecular cloud.

The photometric surface densities are within a factor of $\sim2$ of the total mass
surface density that would be needed to make the dynamical pressure equal to the
H\textsc{ii} region thermal pressure, which is $\Sigma_{\rm tot}=([2/\pi G]P_{\rm
H\textsc{ii}})^{0.5}\sim32\;M_\odot$ pc$^{-2}$. For the H\textsc{ii} region turbulent
pressure of $2.5\times10^5\;k_{\rm B}$, the total mass surface density needed for binding
is $\Sigma_{\rm tot}=85\;M_\odot$ pc$^{-2}$. The dynamical surface density of
$10^3\;M_\odot$ pc$^{-2}$ is more than enough to confine the 
H\textsc{ii} region.

The SFR determined from H$\alpha$ is $\sim0.070\;M_{\odot}$ yr$^{-1}$ for the
$3.6^{\prime\prime}$ square core (430 pc) measured in the head. This gives a rate per
unit area of $\Sigma_{\rm SFR}=0.38\;M_\odot$ pc$^{-2}$ Myr$^{-1}$.  This rate exceeds an
empirically determined limit for the formation of a superwind \citep{heckman02}, which is
$\sim0.1\;M_\odot$ pc$^{-2}$ Myr$^{-1}$, making it plausible that the HII region or hot
gases associated with it are not bound to the galaxy.

\subsubsection{Enhanced Star Formation Rates}

An estimate for the molecular gas surface density may be made from the product of
$\Sigma_{\rm SFR}$ and the typical molecular gas consumption time, which is
$\tau_{gas}\sim2$ Gyr in normal galaxy disks \citep{bigiel08,leroy08} and the outer
regions of spiral galaxies, as obtained by stacking CO spectra when the individual
regions are too faint to see \citep{schruba11}. For $\Sigma_{\rm SFR}=0.38\;M_\odot$
pc$^{-2}$ Myr$^{-1}$ calculated above, the associated molecular surface density would be
$760\;M_\odot$ pc$^{-2}$. This is comparable to the dynamical surface density of
$1.3\times10^3\;M_\odot$ pc$^{-2}$ (Sect. \ref{sect:confine}). Atomic gas would increase this gas surface density even
more. Possibly, the consumption time is shorter than the canonical value; $\tau_{\rm
gas}=66$ Myr would allow $\Sigma_{\rm gas}\sim\Sigma_{\rm stars}\sim25\;M_\odot$
pc$^{-2}$ for the observed SFR surface density. This short consumption time would place
the head of Kiso 5639 a factor of $\sim30$ above the normal Kennicutt-Schmidt
relation between star formation surface density and gas surface density, and in the ULIRG regime \citep[e.g.,][]{sargent14}.

Enhanced SFRs per unit gas mass can also be inferred from our previous observations of 10
XMP galaxies \citep{jorge15}.  There we derived a ratio of $f_{\rm Z}=Z_{\rm
burst}/Z_{\rm host}=0.3-0.1$ for metallicity in the starburst compared to the host, and a
factor of $f_{\rm SFR}=\Sigma_{\rm SFR,burst}/\Sigma_{\rm SFR,host}=10-100$ for the star
formation surface density enhancement. For the present galaxy, Kiso 5639, $f_{\rm Z}=0.4$
and $f_{\rm SFR}=40$ \citep{jorge13}. Simulations by \cite{ceverino16} obtain similar
factors. These factors place constraints on the relative efficiency of star formation in
the burst and on the previous metallicity of the cosmic gas relative to the host, written
as the factor $f_{\rm cosmic}$. We assume that the starburst region contains gas column
density contributions from the cosmic cloud, $\Sigma_{\rm cosmic}$, and from the host,
$\Sigma_{\rm host}$, in some ratio $C$. Then
\begin{equation}
f_{\rm Z}={{f_{\rm cosmic}\Sigma_{\rm cosmic}+\Sigma_{\rm host}}\over{\Sigma_{\rm cosmic}+\Sigma_{\rm host}}} =
{{f_{\rm cosmic}C+1}\over{C+1}}.
\end{equation}
We also assume that the SFR scales with gas column density to a power $\alpha$ but has an
additional efficiency enhancement factor $E$ in the starburst compared to the host. Then
\begin{equation}
f_{\rm SFR}=E(C+1)^\alpha.
\end{equation}
For the Kennicutt-Schmidt relation with total gas, $\alpha\sim1.5$ \citep{kennicutt12}.
Setting $f_{\rm Z}\sim0.4$ and $f_{\rm SFR}\sim40$ as for Kiso 5639 and the galaxies in
\cite{jorge15}, we find that if there is no enhancement in efficiency ($E=1$) then the
accreted gas surface density factor is $C=11$ and the ratio of accreted metallicity to
host metallicity is $f_{\rm cosmic}=0.34$. These numbers decrease as the enhancement
increases: for $E=4$ to 10, $C=3.6$ to 1.5, and $f_{\rm cosmic}=0.24$ to 0.005.  Unless
the accreted gas surface density is significantly higher than the host gas surface
density (i.e., unless $C$ is large), the enhancement above the normal KS relation ($E$)
has to be high and the relative metallicity of the accreted gas ($f_{\rm cosmic}$) has to
be low.

The dense parts of individual molecular clouds \citep{lada10} have relatively short
consumption times like the small values implied by large $E$ (the relative consumption
time is inversely proportional to $E$), but usually observations on large scales like
the $\sim830$ pc head region of Kiso 5639 have the canonical $\sim2$ Gyr and no enhanced
star formation \citep[i.e., $E\sim1$,][]{bigiel08,leroy08}. We conclude that the
accretion of an amount of low metallicity gas comparable to or a few times larger than
the previous gas column density in the disk triggered a substantial burst star formation
with a relative short gas consumption time.

\subsubsection{Head Formation by Gravitational Instability}

The largest star formation complexes in many galaxy disks form by gravitational instabilities
in the turbulent interstellar medium.  We are interested in whether the expected size and mass
of a region formed this way match the tadpole head in Kiso 5639. The characteristic mass for an
instability depends on the gas velocity dispersion and the disk column density. For a
cosmological context with active accretion, \cite{ceverino10} suggest that the clump mass,
$M_{\rm c}$, formed by such an instability is related to the disk mass, $M_{\rm d}$, as $M_{\rm
c}=0.27\delta^2M_{\rm d}$ where $\delta=M_{\rm d}/M_{\rm total}$ and $M_{\rm total}$ is the
total disk mass out to the radius of the clump. They also suggest the clump radius is $R_{\rm
c}\sim0.52\delta R_{\rm d}$.

For Kiso 5639, the head region could be a clump that formed by a gravitational
instability after fresh gas accreted in this part of the galaxy. To compare the
observations with this theoretical prediction, the clump mass is taken to be twice the
total SED mass of the head measured without interclump background subtraction, or
$M_{\rm c}\sim 6.2\times10^6\;M_\odot$. The factor of 2 accounts for gas that we do not
detect. The disk mass is assumed to be half of the dynamical mass because of dark
matter contributing to the rotation curve; then $M_{\rm d}\sim0.75\times10^8\;M_\odot$ and $\delta=0.5$. Thus, the theory in
\cite{ceverino10} predicts that $M_{\rm c}=0.07M_{\rm d}$, and the observations agree,
with $M_{\rm c}\sim0.08M_{\rm d}$. Also for $\delta=0.5$, the theory predicts that the
clump size is 25\% of the galaxy size, and that is also about right for the head.

\subsubsection{Holes and Feedback in the Head H\textsc{ii} region}
\label{holes}

\begin{figure}
\center{\includegraphics[scale=0.45]{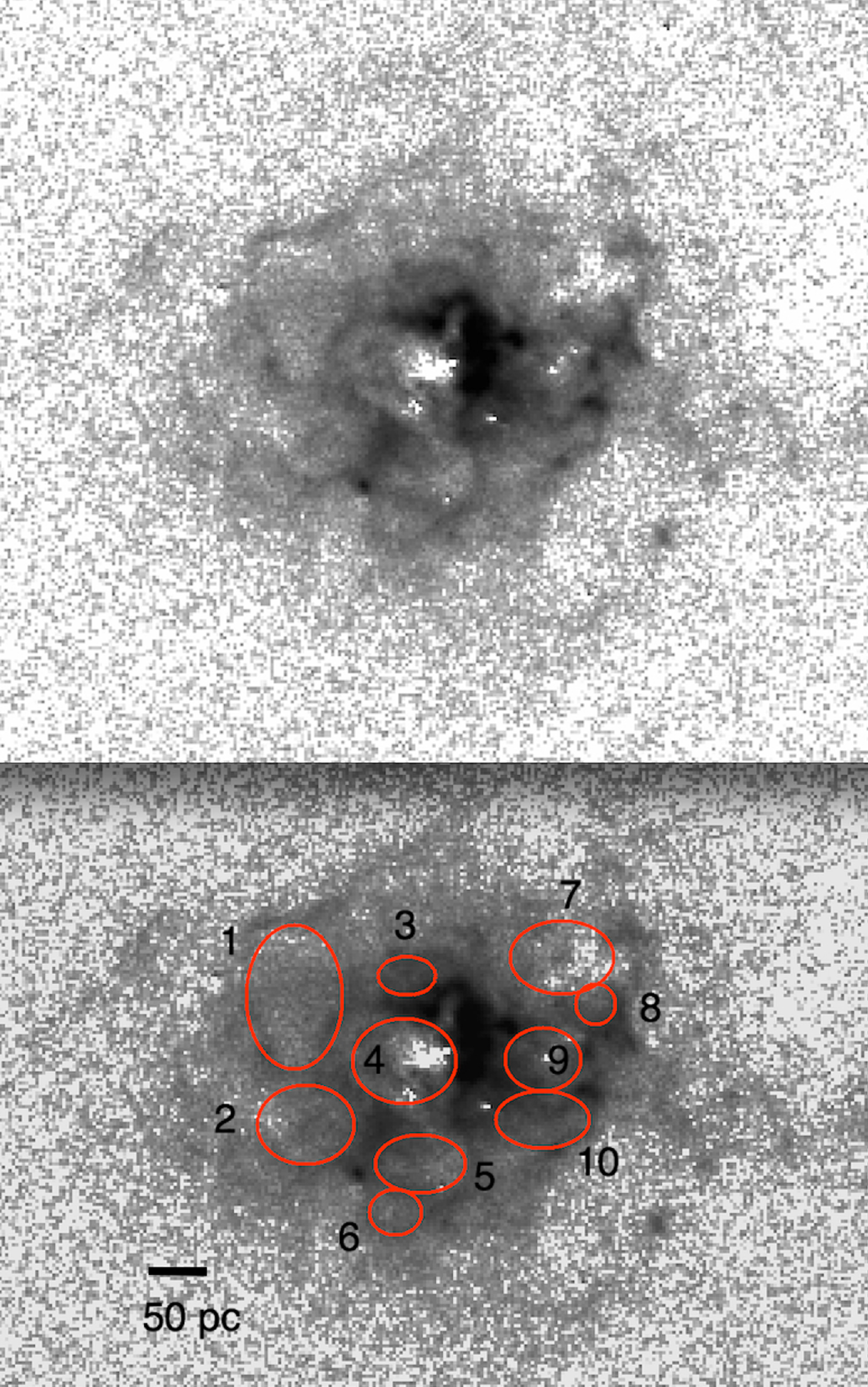}}
\caption{(top) Enlarged logarithmic H$\alpha$ continuum-subtracted image showing the head.
(bottom) Same as top, with ellipses outlining the main holes in the head. Numbers correspond
to the listing in Table \ref{tab3}.}\label{Hahead}\end{figure}

Holes are identified in the continuum-subtracted H$\alpha$ image as low emission regions
surrounded by ring-like structures. The holes evident in Figure \ref{Hahead} (top) are
marked by red ellipses in the bottom panel. They span diameters of 28 to 100 pc, as listed
in Table \ref{tab3}. Hydrodynamic calculations of supergalactic winds show that holes and
filamentary structures can develop in the presence of a high concentration of super star
clusters, such as those observed in M82 \citep{tenorio}. The holes and filaments in Kiso
5639 could have the same origin.

\begin{deluxetable*}{lcccccc}
\tabletypesize{\scriptsize}\tablecolumns{7} \tablewidth{0pt} \tablecaption{H$\alpha$ Holes in Head}

\tablehead{\colhead{Number}&\colhead{RA}&\colhead{Dec}&\colhead {x diameter}&\colhead{y diameter} & \colhead{x diameter}& \colhead{y diameter} \\
\colhead{}&\colhead{(J2000)}&\colhead{J2000}&\colhead{''}&\colhead{''}&\colhead{pc}&\colhead{pc}}

\startdata

1	&11:41:07.700    &+32:25:37.70	&0.57	&0.85	&67.5	&100.9\\
2	&11:41:07.689	&+32:25:36.20	&0.57	&0.47	&67.6	&55.6\\
3	&11:41:07.596	&+32:25:37.95	&0.34	&0.23	&40.0	&27.0\\
4	&11:41:07.598	&+32:25:36.94	&0.61	&0.50	&72.7	&59.4\\
5	&11:41:07.583	&+32:25:35.74	&0.53	&0.34	&63.5	&40.9\\
6	&11:41:07.606	&+32:25:35.17	&0.31	&0.27	&36.9	&32.1\\
7	&11:41:07.451	&+32:25:38.16	&0.61	&0.43	&72.9	&51.2\\
8	&11:41:07.420	&+32:25:37.61	&0.23	&0.24	&27.8	&28.3\\
9	&11:41:07.469	&+32:25:36.97	&0.45	&0.37	&53.4	&44.2\\
10	&11:41:07.469	&+32:25:36.26	&0.55	&0.34	&65.3	&40.2

 \label{tab3}
\enddata

\end{deluxetable*}

The cavities in the head of Kiso 5639 could form by a combination of stellar winds and
supernovae, depending on the age.  Stellar winds would have to dominate if this is a very
young region, as suggested by the photometry. With an ambient density of 1 atom cm$^{-3}$
as calculated above for the H\textsc{ii} region, the wind luminosity $L_{33}$, in units
of $10^{33}$ ergs s$^{-1}$, and the wind duration $t_{\rm My}$, in units of one million
years, that are required to make a wind-swept bubble 50 pc in radius satisfy the product
$L_{33}t_{\rm Myr}^3=2.5\times10^4$, according to equation (6) in \cite{castor75}. An O9
star with solar metallicity has a mass loss rate of ${\dot M}\sim10^{-7.5}$ and wind
speed of $\sim2000$ km s$^{-1}$ \citep{martins15}, so $L_{33}=40$. The mass loss rate
decreases with metallicity as $Z^{0.83}$ \citep{martins15}, where $Z\sim0.1$ for the head
of Kiso 5639, making $L_{33}=6$. Thus, an equivalent of $\sim4000$ O9 stars would be
needed for each large bubble. An O5 star has about 100 times the mass loss rate of an O9
star \citep{martins15}, so 40 equivalent O5 stars would be needed to evacuate the
cavities with stellar winds alone. These numbers of O stars produce between
$1.5\times10^{52}$ and $1.4\times10^{51}$ ionizing photons s$^{-1}$, respectively
\citep{vacca96}, which is consistent with the observed rate of $6.4\times10^{51}$ ph
s$^{-1}$.

Multiple supernovae should help clear the cavities if the age is older than a few Myr. If we
consider that a single supernova deposits a mechanical energy of 1 to 10 $\times10^{49}$ ergs
to the surrounding gas \citep[assuming 1\% to 10\% efficiency,][]{walch15}, and that the
supernova rate is $S$ supernovae per million years, then the average supernova mechanical
luminosity is $3-30\times10^{35}S$ erg s$^{-1}$ or $(300-3000)SL_{33}$. Setting this to
$2.5\times10^4$ for a 1 million year duration in the Castor et al. solution above, we get a
number of supernova $S$ equal to 8-80 to make a 100 pc diameter bubble in that time. For a 3
Myr duration, the number is smaller by $3^3=27$.  These are reasonable numbers if the duration
of star formation is long enough to have had supernovae.

The hole morphology of the head region and the peripheral wispy H$\alpha$ emission are
similar in appearance to the holes and wisps in I Zw 18 \citep[see Figs. 1 and 2
in][]{hunter}. I Zw 18 is one of the lowest metallicity local galaxies \citep{searle,
izotov}. Our estimates for supernovae clearing in Kiso 5639 are consistent with similar
conclusions about star formation in I Zw 18 based on HST WFPC2 images
\citep{hunter,annibali,lebout}.

Supernova clearing of cavities may also lead to gas expulsion from the whole galaxy. Many
of the shells in Figure \ref{Hahead} appear to be broken around the periphery of the
head, suggesting a release of hot gas into the halo. Considering the low escape speed
from this galaxy, $\sim50$ km s$^{-1}$ \citep{jorge13}, and the high thermal speed of a
$10^6$ K supernova cavity, $\sim80$ km s$^{-1}$, localized break out from the head should
mean that a sizeable fraction of the cavity gas escapes from the disk region. This gas
should ultimately become enriched with heavy elements, and if the escaping gas is
preferentially enriched, then the metallicity in the remaining gas and stars can remain
small. This is in accord with current models for the origin of the mass-metallicity
relation in galaxies \citep[e.g.,][]{peeples11,dave12,lilly13,fu13}.

\subsubsection{Clustering Fraction of Star Formation}

There are two indications that a high fraction of star formation in the head of Kiso 5639
is clustered, rather than dispersed. The first was mentioned above, namely that the
summed mass of the measured clusters, $4.0\times10^5\;M_\odot$, produces 30\%-40\%
of the H$\alpha$ flux that is observed.

The second indication comes from the summed cluster masses in the head relative to the
total young stellar mass in the head, which is $10^{6.2\pm0.12}\;M_\odot$ (Table
\ref{tab2}. The ratio is 25\%, but this ignores smaller clusters. We can
integrate over the cluster mass function below the detection limit to include those
smaller clusters. The cluster mass function is $N(M)=N_0M^{-1.73\pm0.51}$ from Section
\ref{masagesfr}. We normalize this to the total count by setting the number greater than
$10^4\;M_\odot$ equal to the observed number of 14. This gives
$N_0=0.73\times14\times10^{4\times0.73}=8.5\times10^3\;M_\odot^{0.73}$. Then the total
integral of cluster mass from a lower limit of $10\;M_\odot$ to $10^4\;M_\odot$ is
$\int_{10}^{10^4} MN_0M^{-1.73}dM=3.2\times10^5\;M_\odot$. This can be added to the
observed mass above $10^4\;M_\odot$, which is $4.0\times10^5\;M_\odot$, to give a total
estimated cluster mass above $10\;M_\odot$ of $7.2\times10^5\;M_\odot$. This total
is 45\% of the young stellar mass.

The high fraction of clustering in the head of Kiso 5639 is inconsistent with the general
lack of clusters in dwarf Irregular galaxies of the same total mass \citep{billett02},
but comparable to the clustering fraction in other starburst dwarfs like NGC 1569 and IC
1705, and in massive active and interacting galaxies \citep{adamo15,adamo15b}.  High
clustering suggests high pressures \citep{kruijssen12}, and highlights the peculiar
nature of the head region studied here.

There are 8 clusters within 50 pc of the center of the head, so the mean separation in 3D
between them is about $100/8^{1/3}=50$ pc. The IRAF task {\it imexamine} was used to
measure the radial profiles of the central clusters to determine their radii at half
maximum intensity. This radius averaged 2 px, which is essentially the resolution limit, so
the maximum diameter of the central clusters is $\sim20$ pc. Thus, the volume filling
fraction of the clusters is $(20/50)^3 = 6$\%. These clusters have a total mass of
$2.5\times10^5 M_\odot$, giving a mass density in the central 50 pc of $0.5\;M_\odot$
pc$^{-3}$, which is equivalent to 14 atoms cm$^{-3}$.

We can compare the density of young clusters in Kiso 5639 with other starburst galaxies. In
the densest clustered region of M82, there are 86 clusters in a 242x242 pc$^2$ area, which
is 1470 clusters per kpc$^2$; in NGC 253, the peak projected cluster density is 113 per
kpc$^2$ \citep{melo}.  In Kiso 5639, the 8 clusters in 50x50 pc$^2$ corresponds to 3200 per
kpc$^2$, which is twice that in M82. So the central region of Kiso 5639 has a high cluster
density. However, the cluster masses in Kiso 5639 are not as large as in M82 where 3
clusters of $\sim6\times10^5\;M_\odot$ each are within 25 pc or each other \citep{west14},
giving a combined mass density of $\sim100\;M_\odot$ pc$^{-3}$.

\subsection{Associated HI gas}
\label{gasmass}

The only H\textsc{i} data available for Kiso 5639 are at low resolution. In an Arecibo
H\textsc{i} survey of star-forming dwarfs, there is a detection near Kiso 5639 with $\log
M(H\textsc{i}) = 8.5$, where the mass is given in $M_{\odot}$ \citep{salzer}. The HIJASS
survey from the Lovell telescope, with a beamsize of 12 arcmin, had a detection in the
region of Kiso 5639 in a common H\textsc{i} halo that included dwarf galaxies Mrk 0746, Was
28, and SDSSJ114135.98+321654.0, with $\log M(H\textsc{i},M_\odot) = 9.2$
\citep{wolfinger}. The total H\textsc{i} mass can be attributed to one, some, or all of
these galaxies. The H\textsc{i} halo shows a small extension to the Northwest, in the
direction of Kiso 5639. There is no optical evidence that these galaxies are interacting.

In the 45 arcsec resolution NRAO VLA Sky Survey (NVSS) by \cite{condon}, there is a 21 cm
radio continuum detection with a flux of 2.7 mJy slightly southwest of the optical position
of Kiso 5639 (the star-forming head). The diffuse nature of the weak emission indicates
that it is likely related to star formation. It is possible that the H\textsc{i} emission
is a reservoir from which accreted gas is stimulating the current star formation.

If the Lyman continuum escape fraction from the star-forming head were comparable to that
estimated for Ly$\alpha$ emitting galaxies, which is 25\% or more \citep{blanc,zheng},
then the ionization rate in the surrounding gas would be $\sim1.6\times10^{51}$
s$^{-1}$, based on the observed rate inside the head of $6.4\times10^{51}$ s$^{-1}$ given
above. The Str{\"o}mgren radius for this ionization could extend to 1 kpc if the average
peripheral density is $0.22$ cm$^{-1}$. This kpc extent would subtend an angle of
$8.4^{\prime\prime}$ in all directions and be easily observed; it is 3/4 of the radius of
the galaxy. The only evidence for peripheral ionization that we see in Figures \ref{HaVB}
and \ref{hag5} is the system of filaments extending away from the body of the galaxy.
Their lengths are typically $1.5-2^{\prime\prime}$ or $\sim$ 150-250 pc.

On the other hand, recent estimates of the Lyman continuum escape fraction for low mass
star-forming galaxies at z$\sim1$ based on HST grism spectroscopy plus GALEX FUV indicate
escape fractions of less than 2\% \citep{rutkow}, which would not be sufficient to
illuminate external gas.

\section{Conclusions}
\label{sect:conc}

HST WFC3 observations of the local metal-poor galaxy Kiso 5639 show a large star-forming
region 830 pc in diameter with 21 clusters of average mass $\sim 2\times 10^4\;
M_{\odot}$ and an H$\alpha$ SFR of $0.070\;M_{\odot}$ yr$^{-1}$. The overall galaxy
appears to be a rotating disk with an inclination of $68^\circ-84^\circ$ (depending on
thickness), but the star-forming region at one end makes it look like a tadpole or comet.
Older and less massive clusters are in four other regions in the tail.

The metallicity in the large starburst region was previously found to be about 40\% of that
in the rest of the galaxy, suggesting that the burst occurred in recently accreted
low-metallicity gas.  A reservoir of HI associated with the galaxy has a mass comparable to
the dynamical mass in the optical disk. This HI could be the source of the accreting gas,
or both the HI and the galaxy disk could be accreting together from a common extragalactic
source.

The properties of the burst region indicate an unusual, high pressure event. The gas
depletion time is shorter than in spiral galaxies by a factor of $5-30$, depending
on the accreting mass, and the clustering fraction of star formation is high, $30-45$\%.
H$\alpha$ observations show 10 cavities in the tadpole head with sizes of 20-100 pc,
which could each arise from many dozens of supernovae over a period of several Myr. The
perimeter of the galaxy shows broken H$\alpha$ shells and filamentary structures that
include some star formation. 

We conclude that the accretion of an amount of low metallicity gas that is comparable to or
a few times larger than the previous gas column density in the disk of the low-metallicity
dwarf galaxy Kiso 5639 triggered a substantial burst of star formation in one localized
region, and that star formation there has a relative short gas consumption time and a high
clustering fraction.

{\it Acknowledgments:} We thank NASA and STScI for observing time and grant support. DME is
supported by HST-GO-13723.002-A and BGE is supported by HST-GO-13723.001-A; both are grateful
to the Severo Ochoa and Jes\'us Serra Foundation for support during a visit to the Instituto de
Astrof\'isica de Canarias. The work of JSA, CMT, and MF has been partly funded by the Spanish
Ministry of Economy and Competitiveness, project AYA2013-47742-C4-2-P. JMA acknowledges support
from the European Research Council Starting Grant (SEDmorph; P.I. V. Wild). MR acknowledges
support from an appointment to the NASA Postdoctoral Program at Goddard Space Flight Center.
DC acknowledges support from the European Research Council Advanced Grant (STARLIGHT; P.I. Ralf Klessen)
This research has made use of the NASA/IPAC Extragalactic Database (NED) which is operated by
the Jet Propulsion Laboratory, California Institute of Technology, under contract with the
National Aeronautics and Space Administration.

\end{document}